%% file: paper.tex
\documentclass[10pt,journal,compsoc]{IEEEtran}
\usepackage{booktabs} 

\usepackage{mathtools, amsfonts, mathrsfs, algorithm2e}
\usepackage[noend]{algorithmic}
\usepackage{makecell, multirow, url}
\usepackage{listings,subcaption}

\usepackage[inline]{enumitem}
\usepackage{xcolor,soul}

\RestyleAlgo{boxruled}
\LinesNumbered

\usepackage{tikz}
\usetikzlibrary{arrows,positioning}
\tikzset{
    >=stealth',
    punkt/.style={
           rectangle,
           rounded corners,
           draw=black, very thick,
           text width=6.5em,
           minimum height=2em,
           text centered},
    pil/.style={
           ->,
           thick,
           shorten <=2pt,
           shorten >=2pt,}
}

\ifCLASSOPTIONcompsoc
  \usepackage[nocompress]{cite}
\else
  \usepackage{cite}
\fi
\ifCLASSINFOpdf
\else
\fi
\begin{document}

%

\title{Enhancing Dynamic Symbolic Execution by Automatically Learning Search Heuristics}

%
%
%
%

\author{Sooyoung Cha, Seongjoon Hong, Jingyoung Kim, Junhee Lee, and Hakjoo
  Oh


\IEEEcompsocitemizethanks{\IEEEcompsocthanksitem S. Cha, S. Hong, J. Kim,
  J. Lee, and H. Oh are with the Department
of Computer Science and Engineering, Korea University, Seoul, Korea.\protect\\
E-mail: hakjoo.oh@gmail.com
}
\thanks{}}

\IEEEtitleabstractindextext{%
  \begin{abstract}
  We present a technique to automatically generate search heuristics
  for dynamic symbolic execution. A key challenge in dynamic symbolic execution is how to
  effectively explore the program's execution paths to achieve high
  code coverage in a limited time budget. Dynamic symbolic execution employs a
  search heuristic to address this challenge, which favors exploring
  particular types of paths that are most likely to maximize the final
  coverage.  However, manually designing a good search heuristic is
  nontrivial and typically ends up with suboptimal and unstable
  outcomes.  The goal of this paper is to overcome this shortcoming of
  dynamic symbolic execution by automatically learning search heuristics.
  We define a class of search heuristics, namely a parametric search heuristic,
  and present an algorithm that efficiently finds an optimal heuristic
  for each subject program. Experimental results with
  industrial-strength symbolic execution tools (e.g., KLEE)
  show that our technique can successfully generate search
  heuristics that significantly outperform existing manually-crafted
  heuristics in terms of branch coverage and bug-finding.
\end{abstract}

\begin{IEEEkeywords}
Dynamic Symbolic Execution, Concolic Testing, Search Heuristics, Software Testing
\end{IEEEkeywords}}

\maketitle

\IEEEdisplaynontitleabstractindextext

%
\IEEEpeerreviewmaketitle


\input{macros}
\input{introduction}
\input{preliminaries}

\input{parametric}
\input{learning}

\input{sec_5-1}
\input{sec_5-2}

\input{sec_5-3}
\input{sec_5-4}
\input{sec_5-5}
\input{relatedwork}
\input{conclusion}
\ifCLASSOPTIONcompsoc
  \section*{Acknowledgments}
\else
  \section*{Acknowledgment}
\fi

This work was supported by Samsung Research Funding \& Incubation Center of Samsung Electronics under Project Number SRFC-IT1701-09.

\ifCLASSOPTIONcaptionsoff
  \newpage
\fi
\bibliographystyle{IEEEtran}
\bibliography{IEEEabrv,paper}

\end{document}

%% file: macros.tex
\newcommand{\mcc}{\mathcal{C}}
\newcommand{\mcg}{\mathcal{G}}
\newcommand{\mcp}{\mathcal{P}}
\newcommand{\mcl}{\mathcal{L}}
\newcommand{\mcu}{\mathcal{U}}
\newcommand{\mcv}{\mathcal{V}}
\newcommand{\mca}{\mathcal{A}}
\newcommand{\mce}{\mathcal{E}}
\newcommand{\mct}{\mathcal{T}}

\newcommand{\myland}{\;\land\;}
\newcommand{\mylor}{\;\vee\;}

\newcommand{\power}[1]{\wp(#1)}

\newcommand{\myset}[1]{\{ #1 \}}
\newcommand{\states}{\it {States}}
\newcommand{\state}{\it {state}}
\newcommand{\testcases}{\it {T}}
\newcommand{\instr}{\it {instr}}
\newcommand{\ninstr}{\it {instr}^\prime}
\newcommand{\store}{\it {S}}
\newcommand{\nstore}{\it {S}^\prime}
\newcommand{\pathcond}{\pi}
\newcommand{\Execute}{\sf {ExecuteInstruction}}
\newcommand{\getCoverage}{\sf {Coverage}}
\newcommand{\paradyse}{\sf {ParaDySE}}

\newcommand{\Oct}{\mbo}
\newcommand{\pre}{\sharp}
\newcommand{\OctPre}{\mbo^{\pre}}
\newcommand{\aoct}{m^{\pre}}
\newcommand{\Var}{\mathit{Var}}
\newcommand{\VPair}{\mathcal{S}}
\newcommand{\pack}{\pi}
\newcommand{\pconf}{\Pi}
\newcommand{\PODom}{\mbo_{\Pi}}

\newcommand{\db}[1]{\llbracket #1 \rrbracket}
\newcommand{\pdb}[1]{\llbracket #1 \rrbracket_{\pconf}}
\newcommand{\ppdb}[1]{\llbracket #1 \rrbracket^{\pre}}
\newcommand{\po}{{po}}

\newcommand{\Classifier}{\mathcal{C}}
\newcommand{\argmax}{\operatornamewithlimits{argmax}}
\newcommand{\todoc}[2]{{\textcolor{#1} {\textbf{[[#2]]}}}}

\newcommand{\todored}[1]{\todoc{red}{#1}}
\newcommand{\todoblue}[1]{\todoc{blue}{#1}}
\newcommand{\todogreen}[1]{\todoc{green}{#1}}

\newcommand*{\TODO}[1]{\todored{#1}}
\newcommand*{\sehun}[1]{\todored{Sehun: #1}}
\newcommand*{\hakjoo}[1]{\todoblue{Hakjoo: #1}}
\newcommand*{\sooyoung}[1]{\todoblue{Sooyoung: #1}}
\newcommand{\commentout}[1]{}

\newcommand{\abst}{}
\newcommand{\aF}{\abst{F}}
\newcommand{\oracle}{{\mathcal{O}}}
\newcommand{\floor}[1]{\lfloor #1 \rfloor}
\newcommand{\mba}{\mathbb{A}}
\newcommand{\mbb}{\mathbb{B}}
\newcommand{\mbc}{\mathbb{C}}
\newcommand{\mbd}{\mathbb{D}}
\newcommand{\mbe}{\mathbb{E}}
\newcommand{\mbf}{\mathbb{F}}
\newcommand{\mbg}{\mathbb{G}}
\newcommand{\mbh}{\mathbb{H}}
\newcommand{\mbi}{\mathbb{I}}
\newcommand{\mbj}{\mathbb{J}}
\newcommand{\mbk}{\mathbb{K}}
\newcommand{\mbl}{\mathbb{L}}
\newcommand{\mbm}{\mathbb{M}}
\newcommand{\mbn}{\mathbb{N}}
\newcommand{\mbo}{\mathbb{O}}
\newcommand{\mbp}{\mathbb{P}}
\newcommand{\mbq}{\mathbb{Q}}
\newcommand{\mbr}{\mathbb{R}}
\newcommand{\mbs}{\mathbb{S}}
\newcommand{\mbt}{\mathbb{T}}
\newcommand{\mbu}{\mathbb{U}}
\newcommand{\mbv}{\mathbb{V}}
\newcommand{\mbw}{\mathbb{W}}
\newcommand{\mbx}{\mathbb{X}}
\newcommand{\mby}{\mathbb{Y}}
\newcommand{\mbz}{\mathbb{Z}}
\newcommand{\cfgto}{\hookrightarrow}

\newcommand*{\cD}{\mathcal{D}}
\newcommand*{\cF}{\mathcal{F}}
\newcommand*{\cP}{\mathcal{P}}
\newcommand*{\cV}{\mathcal{V}}

\newcommand{\mcs}{\mathcal{S}}
\newcommand{\wv}{\mathbf{w}}
\newcommand{\Pgm}{{\it Pgm}}

\newcommand{\Hint}{{\it Profile}}
\newcommand{\hint}{R}
\newcommand{\compo}{\mbj}
\newcommand{\importance}{\mbr}
\newcommand{\feat}{f}
\newcommand{\features}{{\vec{\feat}}}
\newcommand{\strategy}{\mathcal{S}}

\newcommand{\score}{{\it score}}

\newcommand{\mse}{E}

\let\oldvec\vec
\renewcommand{\vec}[1]{\mathbf{#1}}

\newcommand{\obj}{{\it obj}}
\newcommand{\model}{\mathcal{M}}
\newcommand{\counter}{\mathcal{C}}
\newcommand{\lfp}{{\sf lfp}}
\newcommand{\widenop}{\triangledown}
\newcommand{\lub}{{\it lub}}
\newcommand{\glb}{{\it glb}}

\newcommand{\LeastPrecise}{\textsc{NoThld}}
\newcommand{\MostPrecise}{\textsc{FullThld}}
\newcommand{\Selective}{\textsc{Ours}}

\newcommand{\State}{\mbs}
\newcommand{\Action}{\mba}
\newcommand{\actionof}{A}
\newcommand{\policy}{\pi}
\newcommand{\Reward}{\mbr}

\newcommand{\pc}{\Phi}
\newcommand{\bc}{\phi}
\newcommand{\epath}{{\sf path}}
\newcommand{\branch}{\bc}

\newcommand{\symvec}{\vec{M_0}}
\newcommand{\ivec}{\vec{i}}
\newcommand{\RunProgram}{{\sf RunProgram}}
\newcommand{\Choose}{{\sf Choose}}
\newcommand{\NegateAndSolve}{{\sf NegateAndSolve}}
\newcommand{\UNSAT}{{\sf UNSAT}}
\newcommand{\SAT}{{\sf SAT}}
\newcommand{\getModel}{{\sf model}}
\newcommand{\getBranches}{{\sf Branches}}
\newcommand{\concolic}{\mcc}
\newcommand{\symbolic}{\mcs}
\newcommand{\Program}{{\it Program}}
\newcommand{\SearchHeuristic}{{\it Heuristic}}
\newcommand{\ExecutionTree}{{\it ExecutionTree}}
\newcommand{\Branch}{{\it Branch}}
\newcommand{\PathCondition}{{\it PathCond}}
\newcommand{\feature}{\pi}
\newcommand{\featurevector}{\psi}
\newcommand{\param}{\theta}
\newcommand{\family}{\mathcal{H}}
\newcommand{\samplespace}{\mbr}
\newcommand{\comments}[1]{{\color{red}{{\bf #1}}}}

\newcommand{\DFS}{{\sf DFS}}
\newcommand{\RandomState}{{\sf Random}$-${\sf State}}
\newcommand{\CovNew}{{\sf CovNew}}
\newcommand{\MinDistance}{{\sf MinDistance}}
\newcommand{\InstrCount}{{\sf InstrCount}}
\newcommand{\BFS}{{\sf BFS}}
\newcommand{\Depth}{{\sf Depth}}
\newcommand{\QueryCost}{{\sf QueryCost}}
\newcommand{\RandomPath}{{\sf Random}$-${\sf Path}}
\newcommand{\CallPathInstrCount}{{\sf CallPath}$-${\sf InstrCount}}
\newcommand{\RoundRobin}{{\sf RoundRobin}}


%% file: introduction.tex
\section{Introduction}

\IEEEPARstart{D}{ynamic} symbolic
execution~\cite{Sen2005,Godefroid2005,Cadar2005,Cadar2006} has emerged
as an effective software-testing method with diverse
applications~\cite{Avgerinos2014,Christakis2016,Zhang2015,cabfuzz,Stephens2016,
  Chau2017,Hernandez2017,Nassim2018,Sun2018,Sun2019}.  The basic idea
of classical symbolic
execution~\cite{King1975,King,Boyer1975,Howden1977} is to run a
program symbolically, using symbolic values as input and producing
program values represented by symbolic expressions.  Dynamic symbolic execution is
a modern variant of classical symbolic execution, which combines
symbolic and concrete execution to mitigate the inherent limitations
of purely symbolic evaluation of programs
(e.g., handling non-linear
arithmetic, external code).  There are two major flavors of dynamic
symbolic execution~\cite{Cadar2013}, namely concolic
testing~\cite{Sen2005,Godefroid2005} and execution-generated
testing~\cite{Cadar2005,Cadar2006}.  Both approaches have been used in
several tools. For instance, CREST~\cite{Burnim2008} and
SAGE~\cite{Godefroid2008} are well-known concolic testing tools and
KLEE~\cite{Cadar2008} is a representative symbolic executor based on
execution-generated testing.

Search heuristics are a key component of both approaches to dynamic
symbolic execution.  Because of the path-explosion problem, it is
infeasible for dynamic symbolic execution tools to explore all
execution paths of any nontrivial programs.  Instead, they rely on a
search heuristic to maximize code coverage in a limited time budget. A
search heuristic has a criterion and guides symbolic execution to
preferentially explore certain types of execution paths of the subject
program according to its criterion. In concolic testing, for example,
the CFDS (Control-Flow Directed Search) heuristic~\cite{Burnim2008}
prioritizes the execution paths that are close to the uncovered
regions of the program and the CGS (Context-Guided Search)
heuristic~\cite{Seo2014} prefers to explore paths in a new context.
In KLEE, the popular embodiment of execution-generated testing, more than 10 search heuristics are
implemented, one of which is the Depth heuristic that prefers to explore the
paths having the lowest number of executed branches. It is well-known
that choosing a right search heuristic determines the effectiveness of
dynamic symbolic execution in
practice~\cite{Seo2014,Burnim2008,Park2012,cabfuzz,Cadar2008,Li2013,Cha2018,Kim2012,Cadar2013}.

However, designing a good search heuristic is a challenging
task. Manually designing a search heuristic is not only nontrivial but
also likely to deliver sub-optimal and unstable results. As we
demonstrate in~Section~\ref{sec:limit}, no existing search heuristics
consistently achieve good code coverage in practice. For instance, in
concolic testing, the CGS heuristic is arguably the state-of-the-art
and outperforms existing approaches for a number of
programs~\cite{Seo2014}.  However, we found that CGS is sometimes
brittle and inferior even to a random heuristic. Likewise, in
execution-generated testing, the performance of the Depth heuristic
significantly varies depending on the program under test.
Furthermore, existing search heuristics came from a huge amount of
engineering effort and domain expertise. As a result, the difficulty
of coming up with a good search heuristic remains as a major open
challenge in dynamic symbolic execution~\cite{Cadar2013,Roberto18}.

To address this challenge, we present $\paradyse$, a new approach that
automatically generates search heuristics for dynamic symbolic
execution. To this end, we use two key ideas. First, we define a {\em
  parametric search heuristic}, which creates a large class of search
heuristics based on common features of symbolic execution.  The
parametric heuristic reduces the problem of designing a good search
heuristic into a problem of finding a good parameter value. Second, we
present a learning algorithm specialized for dynamic symbolic
execution. The search space that the parametric heuristic poses is
intractably large. Our learning algorithm effectively guides the
search by iteratively refining the search space based on the feedback
from previous runs of dynamic symbolic execution.

Experimental results show that our automatically-generated heuristics
outperform existing manually-crafted heuristics for a
range of C programs. To demonstrate the effectiveness for both flavors of dynamic symbolic
execution, we have implemented $\paradyse$ in CREST~\cite{Burnim2008}
(concolic testing) and KLEE~\cite{Cadar2008} (execution-generated
testing). For the former, we evaluated it on 10 open-source C programs
(0.5--150KLoC). For the latter, we assessed it on the latest versions
of GNU Coreutils. For every benchmark program, our technique has
successfully generated a search heuristic that achieves considerably
higher branch coverage than the existing state-of-the-art
techniques. We also demonstrate that the increased coverage by our
technique leads to more effective finding of real bugs.

\subsection{Contributions}
This paper makes the following contributions:
\begin{itemize}
\item We present $\paradyse$, a new approach for automatically generating
search heuristics for dynamic symbolic execution. Our work represents a
significant departure from prior work; while existing work
(e.g.~\cite{Seo2014,Burnim2008,Park2012,cabfuzz,Cadar2008,Li2013}) focuses on
manually developing a particular search heuristic, our goal is to automate
the very process of generating such a heuristic.

\item We present a parametric search heuristic
  and a learning algorithm for finding good parameter values.

\item We extensively evaluate our approach with open-source C programs.
  We make our tool, called ParaDySE, and data publicly
  available.\footnote{{\underbar{Para}metric \underbar{Dy}namic
      \underbar{S}ymbolic \underbar{E}xecution}: https://github.com/kupl/ParaDySE}
\end{itemize}

This paper is an extension of the previous work~\cite{Cha2018}
presented at 40th International Conference on Software Engineering
(ICSE 2018). The major extensions are as follows:
\begin{itemize}
\item The present paper describes our technique in a generalized
  setting 
and applies it to both approaches (concolic
testing and execution-generated testing) to dynamic
symbolic execution. The use of the previous technique~\cite{Cha2018} was limited to
the concolic testing approach and it was not clear how to apply the idea to
another major approach to dynamic symbolic execution (e.g.,
KLEE~\cite{Cadar2008}).
In Sections~\ref{sec:egt} and~\ref{sec:egtparam}, we explain how to
use our technique to the KLEE-style dynamic symbolic execution.



\item We provide new, extensive experimental results with
  KLEE~\cite{Cadar2008} (Section~\ref{sec:experiments}).

\end{itemize}


%% file: preliminaries.tex
\section{Preliminaries}\label{sec:preliminaries}

In this section, we describe two major approaches to dynamic symbolic
execution, namely concolic testing (Section~\ref{sec:concolic}) and
execution-generated testing (Section~\ref{sec:egt}), and their
limitations (Section~\ref{sec:limit}).

Both concolic testing and execution-generated testing attempt to mix
concrete and symbolic execution. However, they are different in their
specific mechanisms to combine them; that is, concolic testing is
driven by concrete execution and explores each program path one-by-one while execution-generated testing is driven
by symbolic execution and forks the execution whenever encountering a
branch in the program.




\subsection{Concolic Testing}\label{sec:concolic}
Concolic testing is a hybrid software testing technique that combines
concrete and symbolic
execution~\cite{Sen2005,Godefroid2005}. DART~\cite{Sen2005},
CUTE~\cite{Godefroid2005}, SAGE~\cite{Godefroid2008}, and
CREST~\cite{Burnim2008} are well-known concolic testing tools.


\subsubsection{Algorithm}\label{sec:concolicterm}
The idea of concolic testing is to repeat the following process until
a given time budget is exhausted: (1) a program is executed with an
initial input; (2) the exercised path condition is collected during
the concrete execution; and (3) the path condition with one branch
negated is solved with an SMT solver to generate the next input.

Concolic testing begins with executing the subject program $P$ with an
initial input $v_0$.
During the concrete execution, the technique maintains a {\em symbolic
memory state} $S$ and a {\em path condition} $\pc$. The symbolic memory is a mapping
from program variables to symbolic values. It is used to
evaluate the symbolic values of expressions. For instance, when $S$ is $[x \mapsto \alpha, y \mapsto \beta+1]$ (variables $x$ and $y$ are
mapped to symbolic expressions $\alpha$ and $\beta+1$ where $\alpha$
and $\beta$ are symbols), the statement $z := x+y$ transfers the symbolic
memory into $[x \mapsto \alpha, y \mapsto \beta+1, z
\mapsto \alpha+ \beta + 1]$. The path condition represents the
sequence of branches taken during the current execution of the
program.
It is updated whenever a conditional statement ${\it if}(e)$
 is encountered. For instance, when $S=[x \mapsto \alpha]$ and $e = x
 < 1$, the path condition $\pc$ gets updated by $\pc \land (\alpha < 1)$.

Let $\pc = \bc_1 \land
\bc_2 \land \cdots \land \bc_n$ be the path condition that results
from the initial execution. To obtain the next input value,
concolic testing chooses a branch condition $\bc_i$ and generates the
new path condition $\pc'$ as follows:
$\pc' = \bigwedge_{j < i} \bc_j \land \neg \bc_i$.
That is, the new condition $\pc'$ has the same prefix as $\pc$ up to
the $i$-th branch with $\bc_i$ negated, so that input values that satisfy
$\pc'$ drive the program execution to follow the opposite branch of
$\bc_i$. Such concrete input values can be obtained from an SMT
solver. This process is repeated until a fixed testing budget runs
out.

\begin{algorithm}[t]
\caption{Concolic Testing~\label{alg:concolic}}
	\SetKwInOut{Input}{Input}
	\SetKwInOut{Output}{Output}
	\Input{Program $P$, budget $N$, initial input $v_0$}
	\Output{The number of branches covered}

\begin{algorithmic}[1]
\STATE $T \gets \langle \rangle$
\STATE $v \gets v_0$
\FOR{$m=1$ to $N$}
\STATE $\pc_m \gets \RunProgram(P,v)$
\STATE $T \leftarrow T\cdot \pc_m$
        \REPEAT
	\STATE $(\pc, \bc_i) \leftarrow \Choose(T)$\qquad$(\pc = \bc_1
        \land \cdots\land\bc_n$)
        \UNTIL{$\SAT (\bigwedge_{j < i}\bc_j \land \neg \bc_i)$}

        \STATE $v \gets \getModel (\bigwedge_{j < i}\bc_j \land \neg \bc_i)$

\ENDFOR
\RETURN $\lvert \getBranches(T) \rvert$

\end{algorithmic}
\end{algorithm}

Algorithm~\ref{alg:concolic} presents concolic testing
algorithm. The algorithm takes a program $P$, an initial input vector
$v_0$, and a testing budget $N$ (i.e., the number of executions of the
program).  The algorithm maintains the execution tree $T$ of the
program, which is the list of previously explored path conditions. The
execution tree $T$ and input vector $v$ are initially empty and the
initial input vector, respectively (lines 1 and 2).  At line 4, the
program $P$ is executed with the input $v$, resulting in the current
execution path $\pc_m$ explored. The path condition is appended to $T$
(line 5).  In lines 6--8, the algorithm chooses a branch to negate.
The function $\Choose$ first chooses a path condition $\pc$ from $T$,
then selects a branch, i.e., $\bc_i$, from $\pc$. Once a
branch $\bc_i$ is chosen, the algorithm generates the new path
condition $\pc' = \bigwedge_{j < i} \bc_j \land \neg \bc_i$.  If $\pc'$ is satisfiable, the
next input vector is computed (line 9), where $\SAT(\pc)$ returns true
iff $\mbox{$\pc$ is satisfiable}$ and $\getModel(\pc)$ finds an input
vector $v$ which is a model of $\pc$, i.e., $v \models \pc$.
Otherwise, if $\pc'$ is unsatisfiable, the algorithm repeatedly tries
to negate another branch until a satisfiable path condition is found.
This procedure repeats for the given budget $N$ and the final number of covered branches $\lvert \getBranches(T) \lvert$ is returned.

\begin{figure*}[h]
	\centering
    \begin{tabular}{cc}
		\includegraphics[clip, trim={0.8cm 0.4cm 1.0cm 0.9cm}, width=8cm]{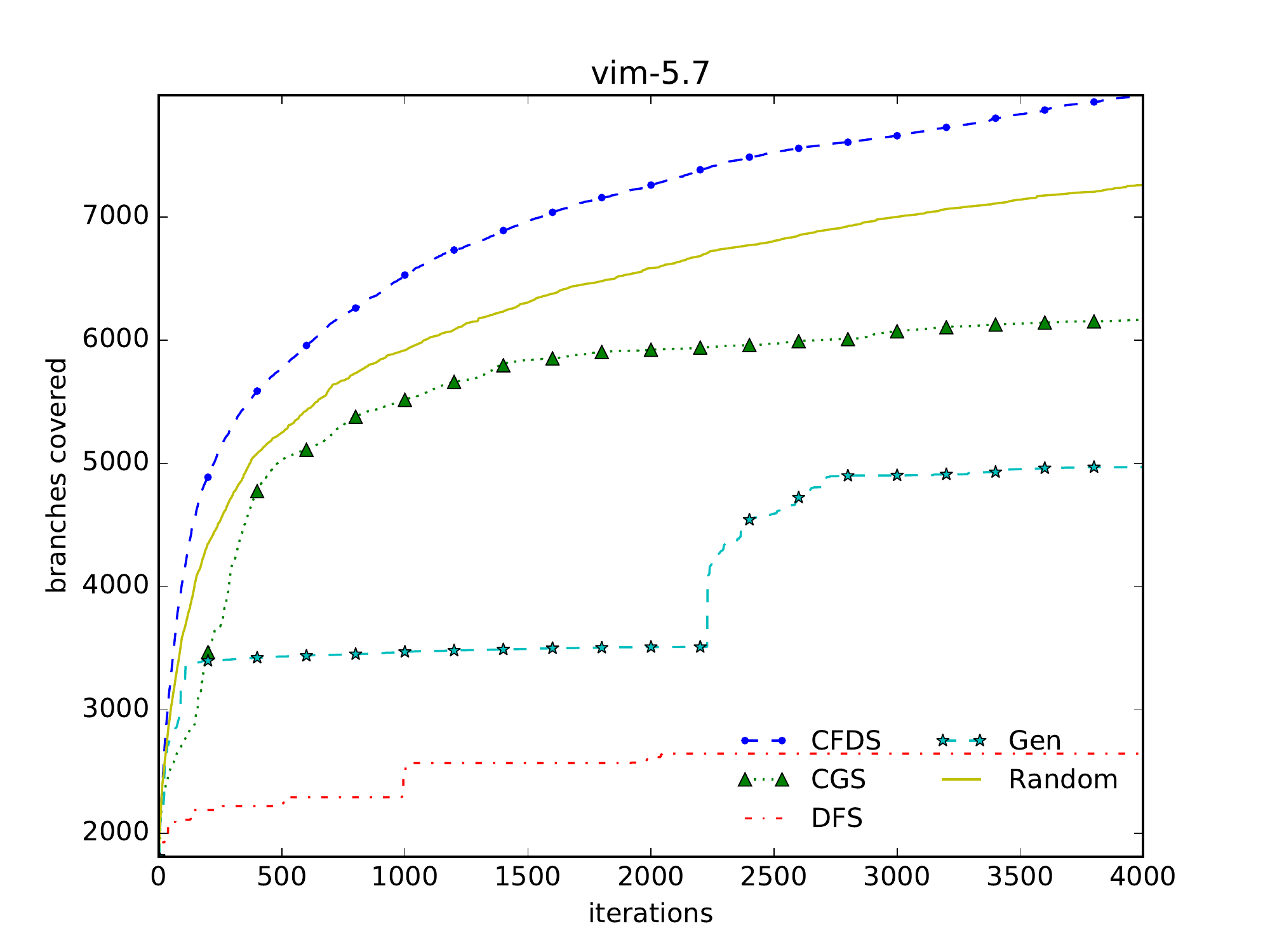} &
		\includegraphics[clip, trim={0.8cm 0.4cm 1.0cm 0.9cm}, width=8cm]{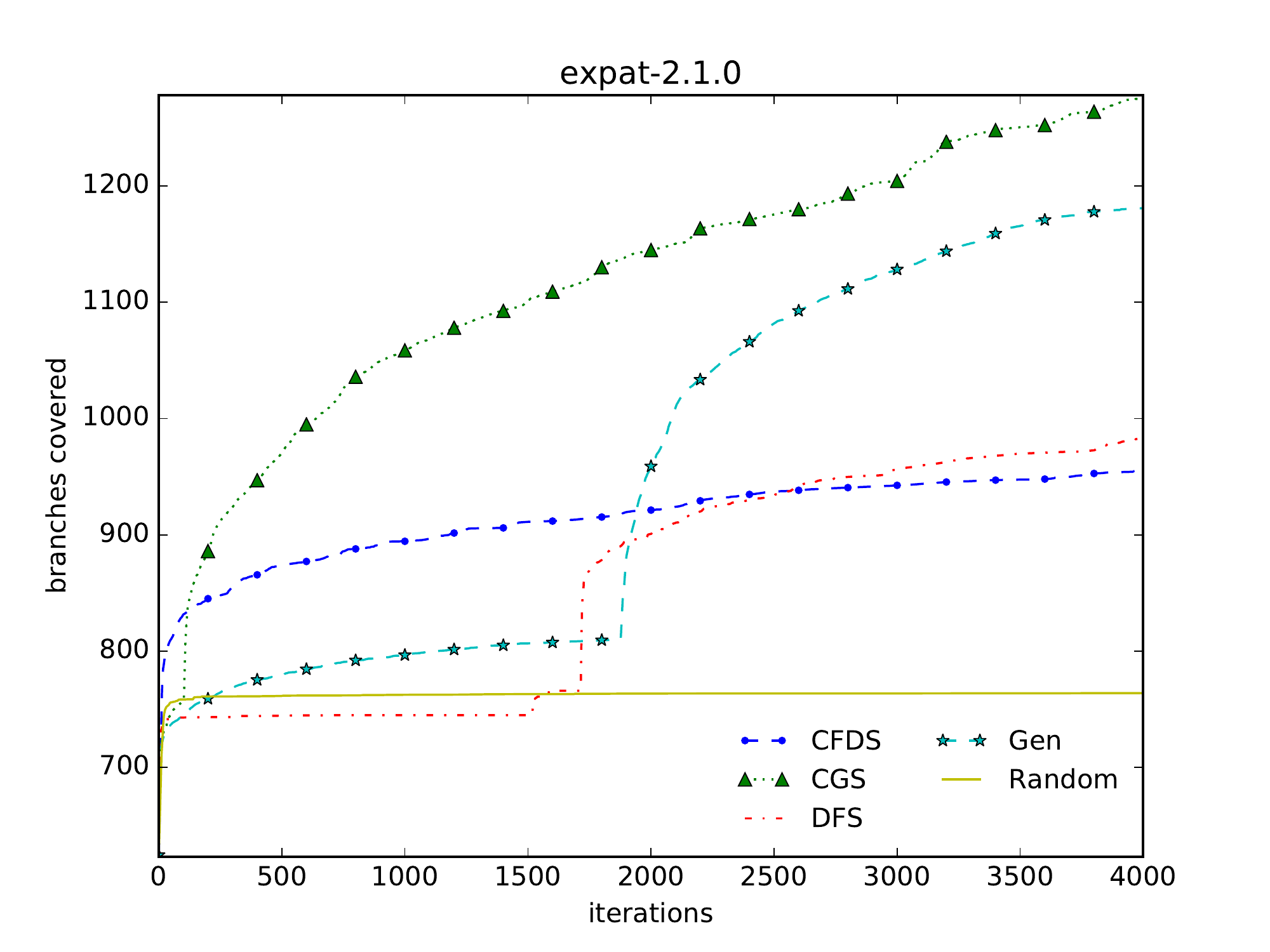}
    \end{tabular}
    \caption{Limitations of existing search heuristics for concolic
      testing: No search heuristic performs well consistently.}
	\label{fig:Moti1}
\end{figure*}

\begin{figure*}[h]
	\centering
    \begin{tabular}{cc}
    	\includegraphics[clip, trim={0.8cm 0.4cm 1.0cm 0.9cm}, width=8cm]{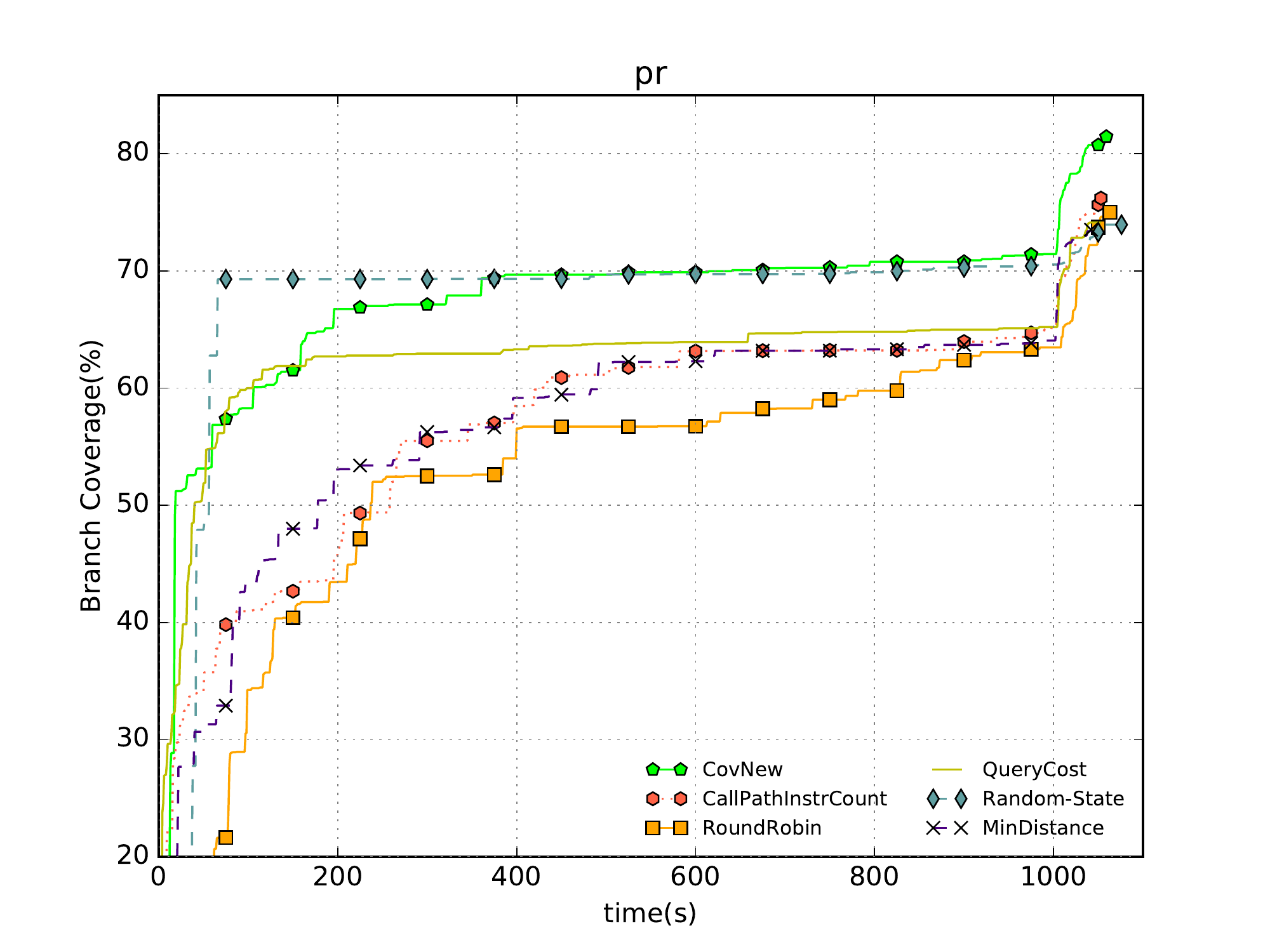} &
		\includegraphics[clip, trim={0.8cm 0.4cm 1.0cm 0.9cm}, width=8cm]{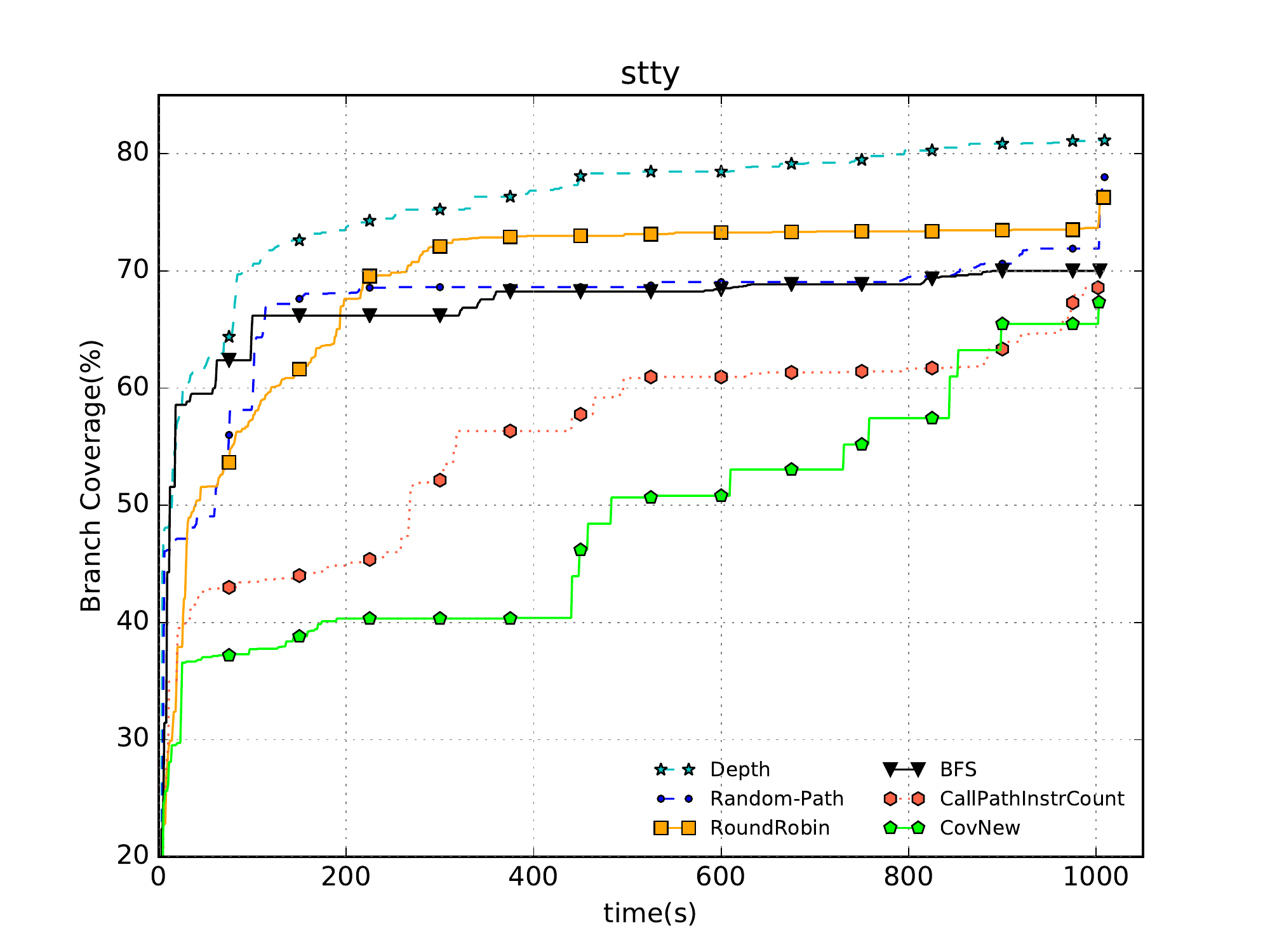}
    \end{tabular}
    \caption{Limitations of existing search heuristics for
      execution-generated testing (KLEE)}
	\label{fig:Moti2}
\end{figure*}

\subsubsection{Search Heuristic}
The performance of Algorithm~\ref{alg:concolic} varies depending on
the choice of the function $\Choose$, namely a search heuristic.
Since the number of execution paths in a program is usually
exponential in the number of branches, exploring all possible
execution paths is infeasible. To address this problem, concolic testing
relies on the search heuristic that steers the execution in
a way to maximize code coverage in a given limited time budget
\cite{Cadar2013}. In prior work, the search heuristic was developed
manually.  Below, we describe three search
heuristics~\cite{Seo2014,Burnim2008}, which are known to perform
comparatively better than others.

The most simple search heuristic, called Random Branch
Search (Random)~\cite{Burnim2008}, is to randomly select a branch from the last
execution path. That is, the $\Choose$ function in
Algorithm~\ref{alg:concolic} is defined as follows:
\[
  \Choose (\langle \pc_1\pc_2\cdots\pc_m \rangle) = (\pc_m,\bc_i)
\]
where $\bc_i$ is a randomly chosen branch from
$\pc_m = \bc_1 \land \cdots \land \bc_n$. Although very simple, the Random
heuristic is typically a better choice than simple deterministic
heuristics such as DFS and BFS~\cite{Burnim2008}.  In our experiments,
the Random heuristic was sometimes better than sophisticated
techniques (Figure~\ref{fig:AC}).

Control-Flow Directed Search (CFDS)~\cite{Burnim2008} is based on the
natural intuition that uncovered branches near the current execution
path would be easier to be exercised in the next execution. This
heuristic first picks the last path condition $\pc_m$, then selects a
branch whose opposite branch is the nearest from any of the unseen
branches.  The distance between two branches is calculated by the
number of branches on the path from the source to the destination. To
calculate the distance, CFDS uses control flow graph of the program,
which is statically constructed before the testing.

Context-Guided Search (CGS)~\cite{Seo2014} basically performs the
breath-first search (BFS) on the execution tree, while reducing the
search space by excluding branches whose ``contexts'' are already
explored.  Given an execution path, the context of a branch in the
path is defined as a sequence of preceding branches.  During search, it
gathers candidate branches at depth $d$ from the execution tree, picks
a branch from the candidates, and the context of the branch is
calculated. If the context has been already considered, CGS skips that
branch and continues to pick the next one. Otherwise, the branch is
negated and the context is recorded. When all the candidate branches
at depth $d$ are considered, the search proceeds to the depth $d+1$ of
the execution tree and repeats the process explained above.


\begin{algorithm}[t]
\caption{Execution-Generated Testing \label{alg:egt}}
	\SetKwInOut{Input}{Input}
	\SetKwInOut{Output}{Output}
	\Input{Program $P$, budget $N$}
	\Output{The number of branches covered}

\begin{algorithmic}[1]
\STATE {$\states \gets \myset{(\instr_{\sf 0}, \store_{\sf 0}, {\it true})}$}
\STATE {$\testcases \gets \emptyset$}
\REPEAT
    \STATE {$(\instr, \store, \pc) \gets \Choose(\states)$}
    \STATE {$\states \gets \states \setminus \myset{(\instr, \store, \pc)}$}
    \STATE {$(\ninstr, \nstore, \pc) \gets \Execute(\myset{\instr, \store, \pc})$}
    \IF {$\ninstr$ = (if ($e$) then $s_1$ else $s_2$)}
        \IF {$\SAT({\it \pc} \land e)$}
            \STATE {$\states \gets \states \cup \myset{(s_1, \nstore, \pc \land e)}$}
        \ENDIF
        \IF {$\SAT({\it \pc} \land \neg e)$}
            \STATE {$\states \gets \states \cup \myset{(s_2, \nstore, \pc \land \neg e)}$}
        \ENDIF
    \ELSIF {$\ninstr$ = halt}
        \STATE {$\testcases \gets \testcases \cup \getModel(\pc)$}
    \ENDIF
    \UNTIL{budget $N$ expires (or $\states = \emptyset$) }
\STATE
\FORALL{$(\instr, \store, \pc) \in \states$}
    \STATE $\testcases \gets \testcases \cup \getModel(\pc)$
\ENDFOR

\RETURN $\lvert \getCoverage(\testcases) \rvert$

\end{algorithmic}
\end{algorithm}

\subsection{Execution-Generated Testing}\label{sec:egt}

Another major flavor of dynamic symbolic execution is
execution-generated testing~\cite{Cadar2005}, which has been implemented in popular symbolic
execution tools such as EXE~\cite{Cadar2006} and
KLEE~\cite{Cadar2008}.

\subsubsection{Algorithm}

Like concolic testing, the main idea of execution-generated testing is
to mix concrete and symbolic execution. Unlike concolic testing,
however, execution-generated testing is not driven by concrete
execution. Instead, it works in a manner similar to pure symbolic
execution and implicitly switches to concrete execution only when path
conditions are unsolvable or the current instruction does not involve
symbolic values.  For our purpose (i.e., focusing on search heuristics), it is sufficient to treat
execution-generated testing as pure symbolic execution.

Execution-generated testing basically maintains a set of states,
where each state is a tuple $(\instr, \store, \pc)$ of
an instruction ($\instr$) to evaluate next, a symbolic memory state
($\store$), and a path condition ($\pc$).\footnote{The definitions of
the symbolic memory state and path condition are given in
Section~\ref{sec:concolicterm}.} Similar to pure symbolic execution,
execution-generated testing forks the execution whenever it evaluates
a conditional statement.


Algorithm~\ref{alg:egt} shows an algorithm for execution-generated testing.
Unlike concolic testing (Algorithm~\ref{alg:concolic}),
the algorithm takes as input a program $P$ and a testing budget $N$
(e.g., 1-hour) only; that is, it does not take the initial input.
At lines 1 and 2, the sets of explored states $\states$ and generated
test-cases $\testcases$ are set to the initial state $(\instr_0,
\store_0, {\it true})$ and the empty set, respectively.
For instance, consider the program:
\begin{lstlisting}
void foo(int x){
  if(x == 20)
    assert("error"); }
\end{lstlisting}
The initial state for the program above is as follows:
\begin{equation}\label{eq:ex1}
({\it if} (x {\it ==} 20), [x \to \alpha], {\it true})
\end{equation}
where the first element represents the next instruction to evaluate,
the symbolic state maps the argument $x$ to symbolic value $\alpha$,
and the path condition is initially ${\it true}$.
At line 4, using the procedure $\Choose$,
the algorithm selects a state to explore from the set $\states$.
At line 6, the algorithm executes the instruction $\instr$ of the
selected state. For simplicity, we consider conditional and halt
instructions and omit other cases, e.g., assignment, assertion. 
If $\ninstr$ is a conditional statement (line 7), the algorithm checks
whether the new path-conditions for the true (line 8)
and false branches (line 10) are satisfiable.
If both conditions are satisfiable, the algorithm forks the state into two states:
$(s_1, \nstore, \pc \land e)$ and $(s_2, \nstore, \pc \land \neg e)$.
For instance, the initial state in~(\ref{eq:ex1}) is split into the states below:
\[
\begin{array}{ccl}
  \state_1 &=& ({\it assert(``error")}, [x \to \alpha], {\it \alpha == 20}) \\
  \state_2 &=& ({\sf halt}, [x \to \alpha], {\it \alpha \neq 20})
\end{array}
\]
When $\ninstr$ is the halt statement (e.g., exit), the algorithm
generates a test-case which is a model of $\pc$ of the state, and then
add it to the set $\testcases$.  The algorithm repeats the process
described above until the time budget $N$ expires or the set $\states$
is empty (line 14). Additionally, at lines 16-18, the algorithm
generates test-cases using the path-conditions of states $\states$,
where the instruction of each state has not yet finished.  Finally,
using the test-cases $\testcases$, the algorithm returns the number of
covered branches (line 18).

\subsubsection{Search Heuristic} Like concolic testing, the
effectiveness of execution-generated testing depends on the choice of
search heuristic, i.e., the $\Choose$ procedure in
Algorithm~\ref{alg:egt}. In this case, $\Choose$ is a function that
takes a set of states and selects a state to explore next. Below, we
describe one representative search heuristic, called RoundRobin, which
is the default search heuristic of KLEE~\cite{Cadar2008} and has been
widely used in prior work (e.g.,~\cite{Cadar2008,Bucur2011,Wong2015}).

The RoundRobin heuristic combines two search heuristics in a round robin
fashion: Random-Path Search (Random-Path) and Coverage-Optimized Search
(CovNew).  The Random-Path heuristic selects a state by randomly traversing
the execution tree on explored instructions of the subject program from the
root. The leaves of the execution tree correspond to the candidate states to
choose from, and the internal nodes denote the locations where the states
forked. Compared to the purely random state selection heuristic (called
Random-State), the Random-Path heuristic prioritizes the states located
higher in the execution tree. The intuition is that the states with fewer
constraints are more likely to reach uncovered code. The CovNew heuristic
first calculates the weights of candidate states and then stochastically
selects the state with high weight. The weight of each candidate is calculated
by two factors; the first one is the minimum distance from the uncovered
instructions, and the second one is the number of executed instructions since
the heuristic most recently covered new instructions.



\subsection{Limitations of Existing Search Heuristics}\label{sec:limit}
Existing search heuristics for both approaches of dynamic symbolic
execution have a key limitation; they rely on a fixed strategy and
fail to consistently perform well on a wide range of target programs.
Our experience with these heuristics is that they are unstable and
their effectiveness varies significantly depending on the target
programs.

Figure~\ref{fig:Moti1} shows that no existing search heuristics
perform consistently in concolic testing.  The branch coverage
achieved by each search heuristic fluctuates with subject programs.
For example, the CFDS heuristic outperforms other existing heuristics
for {\tt vim-5.7} while the heuristic does not perform well for {\tt
  expat-2.1.0}. Conversely, the CGS heuristic achieves the highest
coverage for {\tt expat-2.1.0}, but is inferior even to the random
heuristic for {\tt vim-5.7}. This is not a coincidence. Similar
situations are observed in other programs (see Figure~\ref{fig:AC});
for example, CGS is better than other heuristics for {\tt grep-2.2},
but fails to win the naive random heuristic on {\tt tree-1.6.0}. That
is, the main feature, {\em contexts}, of CGS is not appropriate for
some programs such as {\tt vim-5.7} and {\tt tree-1.6.0}.

For execution-generated testing, we also obtained similar
results. 
We evaluated the 11 search heuristics, including RoundRobin,
implemented in KLEE~\cite{Cadar2008} and Figure~\ref{fig:Moti2} shows
the branch coverage achieved by the top 6 search heuristics for {\tt
  pr} and {\tt stty} in GNU Coreutils-8.31. The CovNew heuristic was
better than other search heuristics for {\tt pr} while the heuristic
took the sixth place for {\tt stty}.  On the other hands, the Depth
heuristic succeeded in achieving the highest branch coverage for {\tt
  stty}, but is not even ranked in the sixth place for {\tt
  pr}. Moreover, as we demonstrate in the experiments
(Figure~\ref{fig:SymbolicAC}), when collecting the search heuristics
with the highest coverage on each of 6 benchmark programs, we obtained
4 distinct heuristics. That is, for each program under test, the most
effective search heuristic is likely to be different.

Besides their sub-optimality, another major limitation of existing
approaches is that developing a good search heuristic requires a huge
amount of engineering effort and expertise. Given that the
effectiveness of dynamic symbolic execution depends heavily on the
search heuristic, ordinary developers who lack the expertise on search
heuristics cannot fully benefit from dynamic symbolic execution. These
observations motivated us to develop a technique that generates search
heuristics automatically.

%% file: parametric.tex
\section{Parametric Search Heuristic}\label{sec:param}
Our first idea for automatically generating search heuristics is to
define a parametric search heuristic, which defines a space of search
heuristics from which our learning algorithm in Section~\ref{sec:algorithm} aims to choose the best
one for each subject program.

In this section, we describe how we parameterize search heuristics
for concolic testing (Section~\ref{sec:concolicparam}) and
execution-generated testing (Section~\ref{sec:egtparam}).
The same idea is used for both approaches with slight variations due
to the different types of search heuristics. 

\subsection{Parameterization for Concolic Testing}\label{sec:concolicparam}
Let $P \in \Program$ be a subject program under test. Recall that a
search heuristic, the $\Choose$ function in
Algorithm~\ref{alg:concolic}, is a function from execution trees to
pairs of a path condition and a branch:
\[
\Choose \in \SearchHeuristic = \ExecutionTree \to \PathCondition \times \Branch
\]
where $\ExecutionTree$ is the set of all execution trees of the program, $\PathCondition$ the set of all path conditions in the trees, $\Branch$ the set of all branches in $P$.

We define a family $\family \subseteq \SearchHeuristic$ of search
heuristics as a parametric heuristic $\Choose_\param$, where
$\theta$ is the parameter which is a $k$-dimensional vector of real
numbers: $\family = \myset{\Choose_\param \mid \param \in \mbr^k}$.
Given an execution tree $T = \langle \pc_1 \pc_2 \cdots \pc_m \rangle$, our parametric search heuristic is defined as follows:
\[
\Choose_{\param}(\langle \pc_1 \cdots \pc_m \rangle) =
(\pc_m,
\argmax_{\bc_j \in \pc_m} \score_\param(\bc_j))
\]

Intuitively, the heuristic first chooses the last path condition $\pc_m$ from the execution tree $T$, then selects a branch $\bc_j$ from $\pc_m$ that gets the highest score among all branches in that path. Except for the CGS heuristic, all existing search heuristics choose a branch from the last path condition. In this work, we follow this common strategy but our method can be generalized to consider the entire execution tree as well. We explain how we score each branch $\bc$ in $\pc_m$ with respect to a given parameter $\param$:

\begin{enumerate}

\item We represent the branch by a feature vector. We designed 40
  boolean features describing properties of branches in concolic testing.
  A feature $\feature_i$ is a boolean predicate on branches:
  \[
  \feature_i : \Branch \to \myset{0,1}.
  \]
For instance, one of the features checks whether the branch is located in the main function or not.
Given a set of $k$ features $\feature = \myset{\feature_1, \dots,
  \feature_k}$, where $k$ is the length of the parameter $\param$, a
branch $\bc$ is represented by a boolean vector as follows:
\[
\feature(\bc) = \langle \feature_1(\bc), \feature_2(\bc),\dots,\feature_k(\bc)  \rangle.
\]

\item 
  Next we compute the score of the branch. In our method, the dimension $k$ of the parameter $\param$ equals to the number of branch features. We use the simple linear combination of the feature vector and the parameter to calculate the branch: 
\[
\score_{\param} (\bc) = \feature(\bc) \cdot \param.
\]

\item Finally, we choose the branch with the highest score. That is, among the branches $\bc_1,\dots,\bc_n$ in $\pc_m$, we choose the branch $\bc_j$ such that $\score_{\param}(\bc_j) \ge \score_{\param}(\bc_k)$ for all $k$.

\end{enumerate}

\input{features}
We have designed 40 features to describe useful properties of branches in concolic testing.
Table~\ref{table:ft} shows the features, which are classified into 12 static and 28 dynamic features.
A static feature describes a branch property that can be extracted without executing the program.
A dynamic feature requires to execute the program and is extracted during concolic testing.


The static features 1-12 describe the syntactic properties of each
branch in the execution path, which can be generated by analyzing the
program text. For instance, feature 8 indicates whether the branch has
a pointer expression in its conditional expression.  We designed these
features to see how much such simple features help to improve branch
coverage, as there is no existing heuristic that extensively considers
the syntactic properties of branches.  At first glance features 2 and
3 seem redundant, but not so.  The true and false branches of loops
have different roles; by giving a high score to a true branch we can
explicitly steer concolic testing away from the loop (i.e. negating
the true branch) while giving a high score to a false branch leads to
getting into the loop.

On the other
hands, 
we designed dynamic features (13-40) to capture the dynamics of
concolic testing. For instance, feature 24 checks whether the branch
has been negated more than 10 times during concolic testing.
That is, during the execution of the program, the boolean value of each dynamic
feature for the same branch may change while the static feature values
of the branch do not.

We also incorporated the key insights of the existing search
heuristics into the features. For example, dynamic features 19-23 were
designed based on the notion of contexts used in the CGS
heuristic~\cite{Seo2014} while features 30-31 are based on the idea of
the CFDS heuristic~\cite{Burnim2008} that calculates the distance to
uncovered branches.

\subsection{Parameterization for Execution-Generated
  Testing}\label{sec:egtparam}

In execution-generated testing, a search heuristic takes 
a set of states and returns a single state. 
Therefore, we define a family $\family \subseteq \SearchHeuristic$ of
search heuristics in this case by the parametric heuristic defined as follows:
\[
\Choose_{\param}(S) = \argmax_{s\in S} \score_\param(s).
\]
The parametric heuristic selects a state $s$ with the highest score
from the set $S$ of states.
Scoring each state with a given parameter $\param$
is similar to the scoring function for concolic testing (Section~\ref{sec:concolicparam}).
The difference is that we need features for describing properties of
states instead of branches of path conditions. The scoring function
$\score_\param$ works as follows: 
\begin{enumerate}
\item It transforms each state in $S$ into a feature vector. We designed 26
  boolean features describing properties of states in execution-generated testing.
  A feature $\feature_i$ is a boolean predicate on states:
  \[
  \feature_i : \states \to \myset{0,1}.
  \]
  With the predefined 26 features, a state $s$ is represented by a boolean vector as follows:
\[
\feature(s) = \langle \feature_1(s), \feature_2(s),\dots,\feature_{26}(s)  \rangle.
\]

\item Second, we compute the state score.
The dimension of the parameter $\param$ is equal to 26,
the number of state features. Using the linear combination,
We calculate the score as follows:
\[
\score_{\param} (\bc) = \feature(\bc) \cdot \param.
\]

\item Finally, we choose from $S$ the state with the highest score.
\end{enumerate}

We have used 26 features to describe useful properties of states in execution-generated testing.
In particular, to reduce the effort for designing the features,
we re-used about half of the 40 branch features
already designed for concolic testing in Table~\ref{table:ft}.
Note that 
the branch features for concolic testing are not immediately available
as state features
for execution-generated testing. 
To reuse the branch features in Table~\ref{table:ft} as state features,
we regarded the reused features as describing the properties of the last branch of the path-condition in each state.
We did not use some features as they are specific to concolic testing.
For instance, we did not use features 28 and 29 in
Table~\ref{table:ft} because there is no
branch negation failure in execution-generated testing. 

Features 1--19 in Table~\ref{table:stateft} show the reused branch features. Specifically,
the features 1--6 belong to the static features in Table~\ref{table:ft}.
For example, feature 5 checks whether the last branch in the path-condition
of each state is true branch of a case statement.
On the other hand, the features 7--19 are dynamic features in Table~\ref{table:ft}.
For instance, feature 10 checks whether the last branch in the path-condition
of each state is selected more than 10 times during execution-generated testing.

We have designed additional 7 state features
to reflect the key insights of the existing search heuristics for execution-generated testing~\cite{Cadar2008}.
The features 20--26 in Table~\ref{table:stateft} represent
the key features of the six relatively effective search heuristics
implemented in KLEE:
Depth, InstrCount, CallPath-InstrCount, QueryCost, MinDistance, and CovNew.
For instance, features 20--21 are based on the idea of the Depth heuristic
that prioritizes the states with the lowest number of executed branches.

\setcounter{rowno}{0}
\begin{table}
\centering
\caption{State features for execution-generated testing.}
\small
\label{table:stateft}
\begin{tabular}{|r|p{7cm}|}
\hline
\# & Description \\
\hline
\rownum & branch in the main function \\
\rownum & true branch of a loop \\
\rownum & false branch of a loop \\
\rownum & branch inside a loop body \\
\rownum & true branch of a case statement \\
\rownum & false branch of a case statement \\
\rownum & branch appearing most frequently \\
\rownum & branch appearing least frequently \\
\rownum & branch located right after the just-selected branch \\
\rownum & branch selected more than 10 times \\
\rownum & branch selected more than 20 times \\
\rownum & branch selected more than 30 times \\
\rownum & branch located in the function of just-selected branch  \\
\rownum & branch is uncovered \\
\rownum & branch selected in the last 10 executions \\
\rownum & branch selected in the last 20 executions \\
\rownum & branch selected in the last 30 executions \\
\rownum & branch in the function that has the largest number of uncovered branches \\
\rownum & branch inside the most recently reached function \\
\hline
\rownum & {10\%} states having the deepest depth \\
\rownum & {10\%} states having the shallowest depth \\
\rownum & {10\%} states with the smallest number of instructions \\
\rownum & {10\%} states with the smallest number of covered instructions in currently executing function \\
\rownum & {10\%} states with the lowest query solving cost \\
\rownum & {10\%} states that are closest to the uncovered instructions \\
\rownum & {10\%} states with the smallest number of executed instructions since the last new instruction was covered. \\

\hline
\end{tabular}
\end{table}

%% file: features.tex
\newcounter{rowno}
\setcounter{rowno}{0}
\newcommand\rownum{\stepcounter{rowno}\arabic{rowno}}

\begin{table}
\centering
\caption{Branch features for concolic testing.
 Features 1--12 are static, and Features 13--40 are dynamic. }
\small
\label{table:ft}
\begin{tabular}{|r|p{7cm}|}
\hline
\# & Description \\
\hline
\rownum & branch in the main function \\
\rownum & true branch of a loop \\
\rownum & false branch of a loop \\
\rownum & nested branch \\
\rownum & branch containing external function calls \\
\rownum & branch containing integer expressions \\
\rownum & branch containing constant strings \\
\rownum & branch containing pointer expressions \\
\rownum & branch containing local variables \\
\rownum & branch inside a loop body \\
\rownum & true branch of a case statement \\
\rownum & false branch of a case statement \\
\hline
\rownum & first {10\%} branches of a path \\
\rownum & last {10\%} branches of a path \\
\rownum & branch appearing most frequently in a path \\
\rownum & branch appearing least frequently in a path \\
\rownum & branch newly covered in the previous execution \\
\rownum & branch located right after the just-negated branch \\
\rownum & branch whose context ($k = 1$) is already visited \\
\rownum & branch whose context ($k = 2$) is already visited\\
\rownum & branch whose context ($k = 3$) is already visited\\
\rownum & branch whose context ($k = 4$) is already visited\\
\rownum & branch whose context ($k = 5$) is already visited\\
\rownum & branch negated more than 10 times \\
\rownum & branch negated more than 20 times \\
\rownum & branch negated more than 30 times \\
\rownum & branch near the just-negated branch  \\
\rownum & branch failed to be negated more than 10 times \\
\rownum & the opposite branch failed to be negated more than 10 times \\
\rownum & the opposite branch is uncovered (depth 0) \\
\rownum & the opposite branch is uncovered (depth 1) \\
\rownum & branch negated in the last 10 executions \\
\rownum & branch negated in the last 20 executions \\
\rownum & branch negated in the last 30 executions \\
\rownum & branch in the function that has the largest number of uncovered branches \\
\rownum & the opposite branch belongs to unreached functions       (top 10\% of the largest func.) \\
\rownum & the opposite branch belongs to unreached functions      (top 20\% of the largest func.) \\
\rownum & the opposite branch belongs to unreached functions      (top 30\% of the largest func.) \\
\rownum & the opposite branch belongs to unreached functions      (\# of branches $>$ 10) \\
\rownum & branch inside the most recently reached function \\
\hline
\end{tabular}
\end{table}


%% file: learning.tex
\section{Parameter Optimization Algorithm} \label{sec:algorithm}
Now we describe our algorithm for finding a good parameter value of
the parametric search heuristic in Section~\ref{sec:param}.
We define the optimization problem, and then present our algorithm.
Our optimization algorithm is general and can be used for both approaches
to dynamic symbolic execution.

\subsection{Optimization Problem}\label{sec:opt-problem}

In our approach, finding a good search heuristic corresponds to
solving an optimization problem.
We model dynamic symbolic execution algorithms (i.e.,
Algorithm~\ref{alg:concolic} and Algorithm~\ref{alg:egt})
by the function:
\[
  \symbolic: \Program \times \SearchHeuristic \to \mbn
\]
which takes a program and a search heuristic, and returns the number
of covered branches.
Given a program $P$ and a search heuristic
$\Choose$, $\symbolic(P,\Choose)$ performs dynamic symbolic execution
using the heuristic for a fixed testing budget (i.e. $N$).
We assume that the initial input ($v_0$) for concolic testing
and the testing budget ($N$) are fixed for each subject program.

Given a program $P$ to test, our goal is to find a parameter $\param$
that maximizes the performance of $\symbolic$ with
respect to $P$.  Formally, our objective is to find $\theta^*$ such
that
\begin{equation}\label{eq:opt}
  \theta^* = \argmax_{\param \in \mbr^{k}} \symbolic(P, \Choose_{\param}).
\end{equation}
That is, we aim to find a parameter $\param^*$ that causes the function
$\symbolic$ with the search heuristic
$\Choose_\param$ to maximize the number of covered branches in $P$.

\subsection{Optimization Algorithm}\label{sec:algorithm}

We propose an algorithm that efficiently solves the optimization
problem in~(\ref{eq:opt}).  A simplistic approach to solve the problem
would be the random sampling method defined as follows:
 \begin{algorithmic}[1]
   \REPEAT \STATE $\param \gets $ sample from $\mbr^k$ \STATE
   $B \gets \symbolic (P, \Choose_\param)$ \UNTIL timeout \RETURN best
   $\param$ found
 \end{algorithmic}
which randomly samples parameter values and returns the best parameter
found for a given time budget. However, we found that this naive
algorithm is extremely inefficient and leads to a failure when it is
used for finding a good search heuristic (Section~\ref{sec:opt-algo}).
This is mainly because of two
reasons. First, the search space is intractably large and therefore
blindly searching for good parameters without any guidance is
hopeless. Second, a single evaluation of a parameter value is
generally unreliable and does not represent the average performance
in dynamic symbolic execution. For example,
the performance of concolic testing can vary due to the inherent
nondeterminism (e.g. branch prediction failure)~\cite{Godefroid2005}.

In response, we designed an optimization algorithm
(Algorithm~\ref{alg:learning}) specialized to efficiently finding good
parameter values of search heuristics.  The key idea behind this
algorithm is to iteratively refine the sample space based on the
feedback from previous runs of dynamic symbolic execution.  The main loop of the
algorithm consists of the three phases: {\em Find}, {\em Check}, and
{\em Refine}. These three steps are repeated until the average
performance converges.

At line 2, the algorithm initializes the sample spaces. It maintains
$k$ sample spaces, $\mbr_i~(i \in [1,k])$, where $k$ is the dimension
of the parameters (i.e., the number of features in our
parametric heuristic).  In our algorithm, the $i$-th components of
the parameters are sampled from $\mbr_i$, independently from other
components.  For all $i$, $\mbr_i$ is initialized to the space
$[-1,1]$.

In the first phase ({\em Find}), we randomly sample $n$ parameter
values: $\param_1, \param_2, \dots, \param_n$ from the current sample
space $\mbr_1 \times \mbr_2\times \cdots \times \mbr_k$ (line 7), and
their performance numbers (i.e., the number of branches covered) are
evaluated (lines 9--11). In experiments,
we set $n$ depending on the given program $P$ (Table~\ref{tb:time}).
Among the $n$ parameters, we choose the top $K$ parameters
according to their branch coverage.  In our experiments, $K$ is set to
10 because we observed that parameters with good qualities are usually
found in the top 10 parameters. This first step of performing
the symbolic execution function $n$ times (line 11) can be run in parallel.

\begin{algorithm}[t]
\small
\caption{Parameter Optimization Algorithm\label{alg:learning}}
    \SetKwInOut{Input}{Input}
    \SetKwInOut{Output}{Output}
    \Input{Program $P$}
    \Output{Optimal parameter $\param \in \mbr^k$ for $P$}
\begin{algorithmic}[1]
{\tiny}
\STATE /* $k$: the dimension of $\param$ */
    \STATE initialize the sample spaces $\samplespace_i=[-1, 1]$ for $i \in [1,k]$
    \STATE $\langle max, \mbox{converge} \rangle \gets \langle 0, {\it false} \rangle$
        \REPEAT
\STATE {/* Step 1: Find */}
        \STATE /* sample $n$ parameters: $\param_1,\dots,\param_n$ (e.g., n=1,000) */
           \STATE $\myset{\param_i}_{i=1}^n$ $\gets$ sample from $\samplespace_1\times\samplespace_2\times\cdots\times\samplespace_k$
           \STATE /* evaluate the sampled parameters */
            \FOR{$i = 1$ \TO $n$}
            \STATE /* $B_i$: branch coverage achieved with $\param_i$ */
            \STATE $B_i \gets$ $\symbolic(P, \Choose_{\param_i}$)
            \ENDFOR
            \STATE pick top $K$ parameters $\myset{\param_i'}_{i=1}^K$ from $\myset{\param_i}_{i=1}^n$ with highest $B_i$
\STATE
\STATE {/* Step 2: Check */}
            \FORALL{$K$ parameters $\param_i'$}
                \STATE $B_i^* \gets \frac{1}{10}\sum_{j=1}^{10} \symbolic(P, \Choose_{\param_i'}$)
            \ENDFOR
            \STATE pick top 2 parameters $\param_{t_1}, \param_{t_2}$ with highest $B_i^*$
\STATE
\STATE {/* Step 3: Refine */}
             \FOR{$i = 1$ \TO $k$}
                \IF{ $\param_{t_1}^i > 0$ \AND $\param_{t_2}^i > 0$ }
                \STATE $\samplespace_i$ = [$\min(\param_{t_1}^i,\param_{t2}^i$), 1]
                \ELSIF{ $\param_{t_1}^i < 0$ \AND $\param_{t_2}^i < 0$ }
                \STATE $\samplespace_i$ = [-1, $\max(\param_{t_1}^i,\param_{t_2}^i$)]
                \ENDIF
            \ENDFOR
\STATE
            \STATE {/* Check Convergence */}
             \IF{$B_{t_1}^* < max$}
             \STATE{converge $\gets$ {\it true}}
             \ELSE
             \STATE {$\langle max, \param_{max} \rangle \gets \langle B_{t_1}^*, \param_{t_1} \rangle$}
             \ENDIF
        \UNTIL converge
        \RETURN $\param_{max}$
\end{algorithmic}
\end{algorithm}

In the next phase ({\em Check}), we choose the top 2 parameters that show
the best average performance. At lines 16--17, the $K$ parameters
chosen from the first phase are evaluated again to obtain the average
code coverage over 10 trials, where $B_i^*$ represents the average
performance of parameter $\param_i'$. At line 19, we choose two
parameters $\param_{t_1}$ (top 1) and $\param_{t_2}$ (top 2) with the best average
performance.  This step ({\em Check}) is needed to rule out unreliable
parameters. Because of the nondeterminism of dynamic symbolic execution, the
quality of a search heuristic must be evaluated over multiple
executions.

In the third step ({\em Refine}), we refine the sample spaces
$\mbr_1,\dots,\mbr_k$ based on $\param_{t_1}$
and $\param_{t_2}$. Each $\mbr_i$ is refined based on the values of the
$i$-th components ($\param_{t_1}^i$ and $\param_{t_2}^i$) of $\param_{t_1}$ and
$\param_{t_2}$. When both $\param_{t_1}^i$ and $\param_{t_2}^i$ are positive, we
modify $\mbr_i$ by $[\min(\param_{t_1}^i, \param_{t_2}^i), 1]$.  When both
$\param_{t_1}^i$ and $\param_{t_2}^i$ are negative, $\mbr_i$ is refined by
$[-1,\max(\param_{t_1}^i, \param_{t_2}^i)]$. Otherwise, $\mbr_i$ remains the
same. Then, our algorithm goes back to the first phase ({\em Find})
and randomly samples $n$ parameter values from the refined space.

Finally, our algorithm terminates when the best average coverage
($B_{t_1}^* $) obtained in the current iteration is less than the coverage
($max$) from the previous iteration (lines 30--31). This way, we iteratively
refine each sample space $\mbr_i$ and guide the search to continuously find
and climb the hills toward top in the parameter space.

%% file: sec_5-1.tex
\section{Experiments}
\label{sec:experiments}

In this section, we experimentally evaluate our approach for
automatically generating search heuristics of dynamic symbolic
execution.
We demonstrate the effectiveness of our approach for both approaches
to dynamic symbolic symbolic execution: concolic testing and
execution-generated testing.
For the former, we implemented our approach, called $\paradyse$, 
in CREST~\cite{crest},
a concolic testing tool widely used for C programs~\cite{Burnim2008,Seo2014,Garg2013,Kim2012}.
For the latter, we implemented $\paradyse$ in KLEE~\cite{Cadar2008},
one of the most popular symbolic execution tools widely used in previous work~\cite{Chipounov2012,Marinescu2012,Yi2015,Yi2018,Kuznetsov2012,Kim2012,Trabish2018}.
We conducted experiments to answer the following research questions:

\begin{itemize}
\item {\bf Effectiveness of generated heuristics}: Does our approach
  generate effective search heuristics for dynamic symbolic execution?
  How do they perform compared to the existing state-of-the-art heuristics?
\item {\bf Time for obtaining the heuristics}: How long does our
  approach take to generate the search heuristics? Is our approach
  useful even considering the training effort?
\item {\bf Efficacy of optimization algorithm}:
  How does our optimization algorithm perform compared to the naive
  algorithm by random sampling?
\item {\bf Important features}: What are the important features to
  generate effective search heuristics for both approaches of dynamic
  symbolic execution, respectively?
\end{itemize}
\noindent
All experiments were done on
a linux machine with two Intel Xeon Processor E5-2630 and 192GB RAM.

\subsection{Evaluation Setting}
\subsubsection{Concolic Testing}
We have compared our approach with five existing heuristics: CGS
(Context-Guided Search)~\cite{Seo2014}, CFDS (Control-Flow Directed
Search)~\cite{Burnim2008}, Random (Random Branch Search)~\cite{Burnim2008}, DFS
(Depth-First Search)~\cite{Godefroid2005}, and Gen (Generational
Search)~\cite{Godefroid2008}. We chose these heuristics for comparison
because they have been commonly used in prior
work~\cite{Seo2014,Burnim2008,Daca2016,Godefroid2005,Godefroid2008}.  In
particular, CGS and CFDS are arguably the state-of-the-art search
heuristics that often perform the best in practice~\cite{Seo2014,Burnim2008}.
The implementation of CFDS, Random, and DFS heuristics are available in CREST.  The
implementations of CGS and Gen came from the prior
work~\cite{Seo2014}.\footnote{We obtained the implementation from authors via personal communication.}

\begin{table}[]
\centering
\caption{Benchmark programs for concolic testing}
\label{table:bench}
\scalebox{0.9}{
\begin{tabular}{@{}lrrc@{}}
\toprule
Program     & \multicolumn{1}{l}{\# Total branches} & \multicolumn{1}{l}{LOC} & \multicolumn{1}{l}{Source} \\ \midrule
vim-5.7     & 35,464                                & 165K                    & \cite{Burnim2008}          \\
gawk-3.0.3  & 8,038                                 & 30K                     & ours                       \\
expat-2.1.0 & 8,500                                 & 49K                     & \cite{Seo2014}\\
grep-2.2    & 3,836                                 & 15K                     & \cite{Burnim2008}           \\
sed-1.17    & 2,656                                 & 9K                      & \cite{Kim2011}             \\
tree-1.6.0  & 1,438                                 & 4K                      & ours                       \\ \midrule
cdaudio     & 358                                   & 3K                      & \cite{Seo2014}                            \\
floppy      & 268                                   & 2K                    & \cite{Seo2014}                           \\
kbfiltr     & 204                                   & 1K                      & \cite{Seo2014}                           \\
replace     & 196                                   & 0.5K                    & \cite{Burnim2008} \\ \bottomrule
\end{tabular}}
\end{table}

\begin{table}[]
\centering
\caption{Benchmark programs for execution-generated testing (KLEE):
  the 6 largest (excluding ones on which KLEE does not run) benchmark
  programs from Coreutils-8.31. }
\label{table:kleebench}
\begin{tabular}{@{}lrrc@{}}
\toprule
Program     & \multicolumn{1}{l}{\# Total branches} & \multicolumn{1}{l}{LOC} \\\midrule
ls          & 1,590                                 & 5K                     \\
dd          & 690                                   & 3K                     \\
pr          & 645                                   & 3K                     \\
ptx         & 577                                   & 2K                     \\
factor      & 445                                   & 3K                     \\
stty        & 420                                   & 2K                     \\\bottomrule
\end{tabular}
\end{table}

We used 10 open-source benchmark programs
(Table~\ref{table:bench}). 
The benchmarks are divided into the large and small programs. The
large benchmarks include
 \texttt{vim},
\texttt{expat}, \texttt{grep}, \texttt{sed},
\texttt{gawk}, and \texttt{tree}.
The first four are standard benchmark programs in concolic testing for
C, which have been used multiple times in prior work~\cite{Boonstoppel2008,
  Burnim2008, Cadar2008, Kim2011, Seo2014}. The last two programs
(\texttt{gawk} and \texttt{tree}) were prepared by ourselves, which
are available with our tool. Our benchmark set also includes 4
small ones: \texttt{cdaudio}, \texttt{floppy}, \texttt{kbfiltr}, and \texttt{replace},
which were used in~\cite{Burnim2008, Kim2011, Seo2014}.




We conducted all experiments under the same evaluation setting; the initial
input (i.e. $v_0$ in Algorithm~\ref{alg:concolic}) was fixed for each
benchmark program and a single run of concolic testing used the same testing
budget (4,000 executions, i.e., $N=4000$ in Algorithm~\ref{alg:concolic}). We
set the budget to 4,000 program executions for fair comparison with
existing techniques; it is the number that has been
commonly used in prior work on search heuristics~\cite{Burnim2008,Seo2014}. Note that the
performance of concolic testing generally depends on the initial input. We
found that in our benchmark programs, except for {\tt grep} and {\tt expat},
different choices of initial input did not much affect the final performance,
so we generated random inputs for those programs. For {\tt grep} and {\tt
expat}, the performance of concolic testing varied significantly depending on
the initial input. 
For
instance, with some initial inputs, CFDS and Random covered 150 less branches
in \texttt{grep} than with other inputs. We avoided this exceptional case
when selecting the input for \texttt{grep} and {\tt expat}. For {\tt expat},
we chose the input used in prior work~\cite{Seo2014}. For {\tt grep}, we
selected an input value on which the Random heuristic was reasonably effective. The initial
input values we used are available with our tool.

The performance of each search heuristic was averaged over multiple
trials.  Even with the same initial input, the search heuristics
have coverage variations for several reasons: search
initialization in concolic testing~\cite{Godefroid2005}, the
randomness of search heuristics, and so on. We repeated the
experiments 100 times for all benchmarks except for \texttt{vim}
for which we averaged over 50 trials as its execution
takes much longer time. 

\subsubsection{Execution-Generated Testing}
We have compared our approach with 11 existing search heuristics
implemented in KLEE~\cite{Cadar2008}:
$\DFS$ (Depth-First Search), $\BFS$ (Breath-First Search),
$\RandomState$, $\RandomPath$, $\CovNew$, $\QueryCost$, $\MinDistance$ (Minimal-Distance to Uncovered),
$\Depth$, $\InstrCount$ (Instruction-Count), $\CallPathInstrCount$ (CallPath-Instruction-Count),
and $\RoundRobin$ using $\RandomPath$ and $\CovNew$ in a round robin fashion.

We used the six largest benchmark programs in GNU Coreutils-8.31
(Table~\ref{table:kleebench}), excluding programs on which KLEE is unable
to run. 
We used GNU Coreutils as it is the most commonly used benchmark for
evaluating KLEE (e.g.,~\cite{Marinescu2012, Kuznetsov2012, Li2013, Yi2015, Yi2018, Wong2015, Mechtaev2018fse}).
The six benchmarks include \texttt{ls}, \texttt{dd}, \texttt{pr}, \texttt{ptx},
\texttt{factor}, and \texttt{stty}.
We excluded small programs (e.g., \texttt{cat}, \texttt{rm}, and
\texttt{pwd}) in Coreutils. The existing search heuristics already achieve high branch coverage on those programs,
as their sizes are quite small (e.g., \texttt{pwd} is of 0.4KLoC).

We used the same evaluation settings in all experiments.  First, we
allocated 1,000 seconds for testing budget (i.e., $N=1000s$ in
Algorithm~\ref{alg:egt}).  Unlike concolic testing, we used the
maximum running time as the testing budget instead of the number of program executions.  This is
because using timing budget has been more popular in previous works on KLEE~\cite{Cadar2008,
  Kim2012, Marinescu2012, Li2013}.  Second, because of the randomness
of search heuristics, we repeated the experiments for all benchmarks
10 times and reported the average branch coverage over 10 trials.


%% file: sec_5-2.tex
\begin{figure*}[h]
	\centering
	\begin{tabular}{cc}
		\includegraphics[clip, trim={0.8cm 0.4cm 1.0cm 0.5cm}, width=8cm]{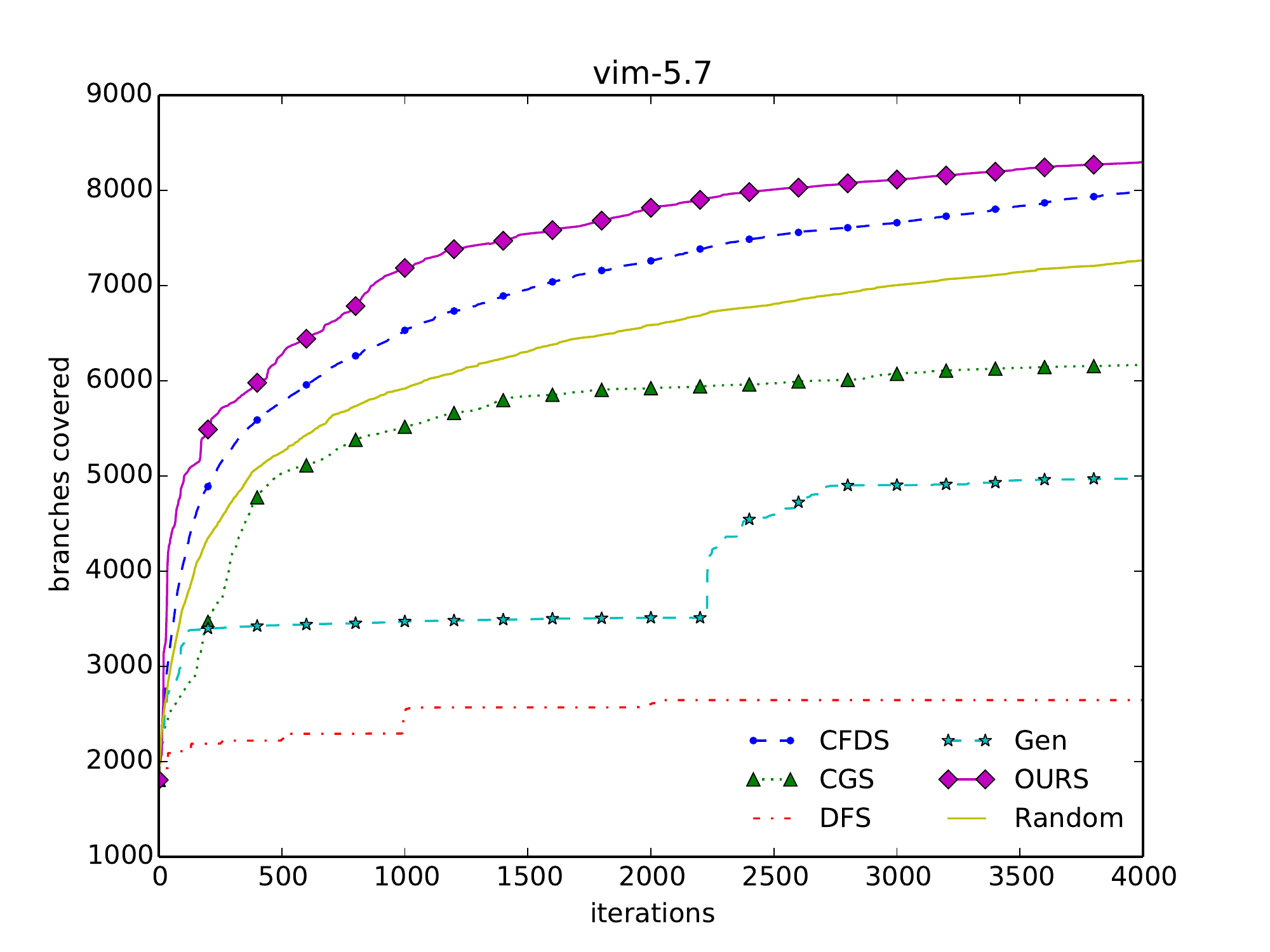} &
		\includegraphics[clip, trim={0.8cm 0.4cm 1.0cm 0.5cm}, width=8cm]{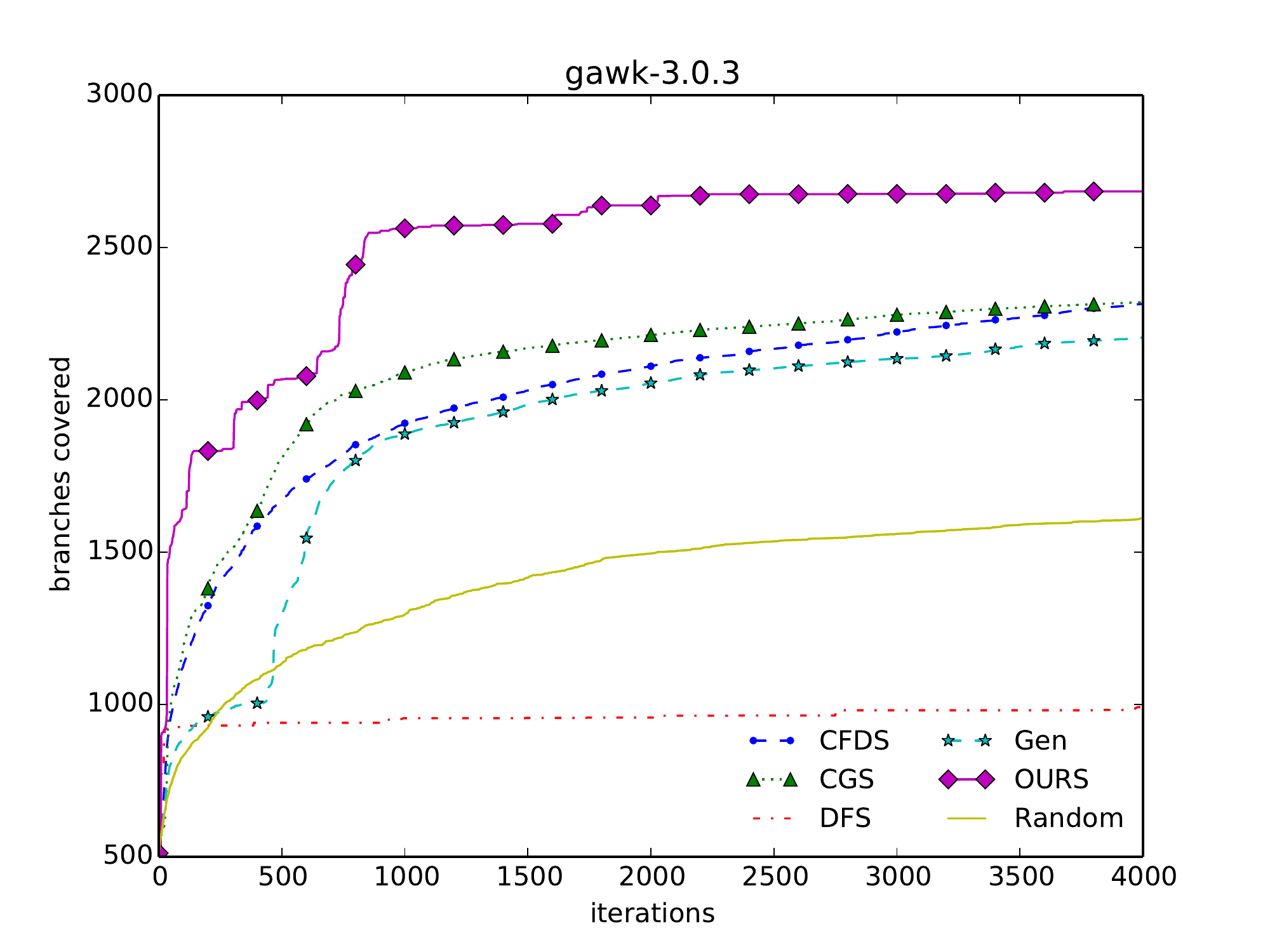} \\
		\includegraphics[clip, trim={0.8cm 0.4cm 1.0cm 0.5cm}, width=8cm]{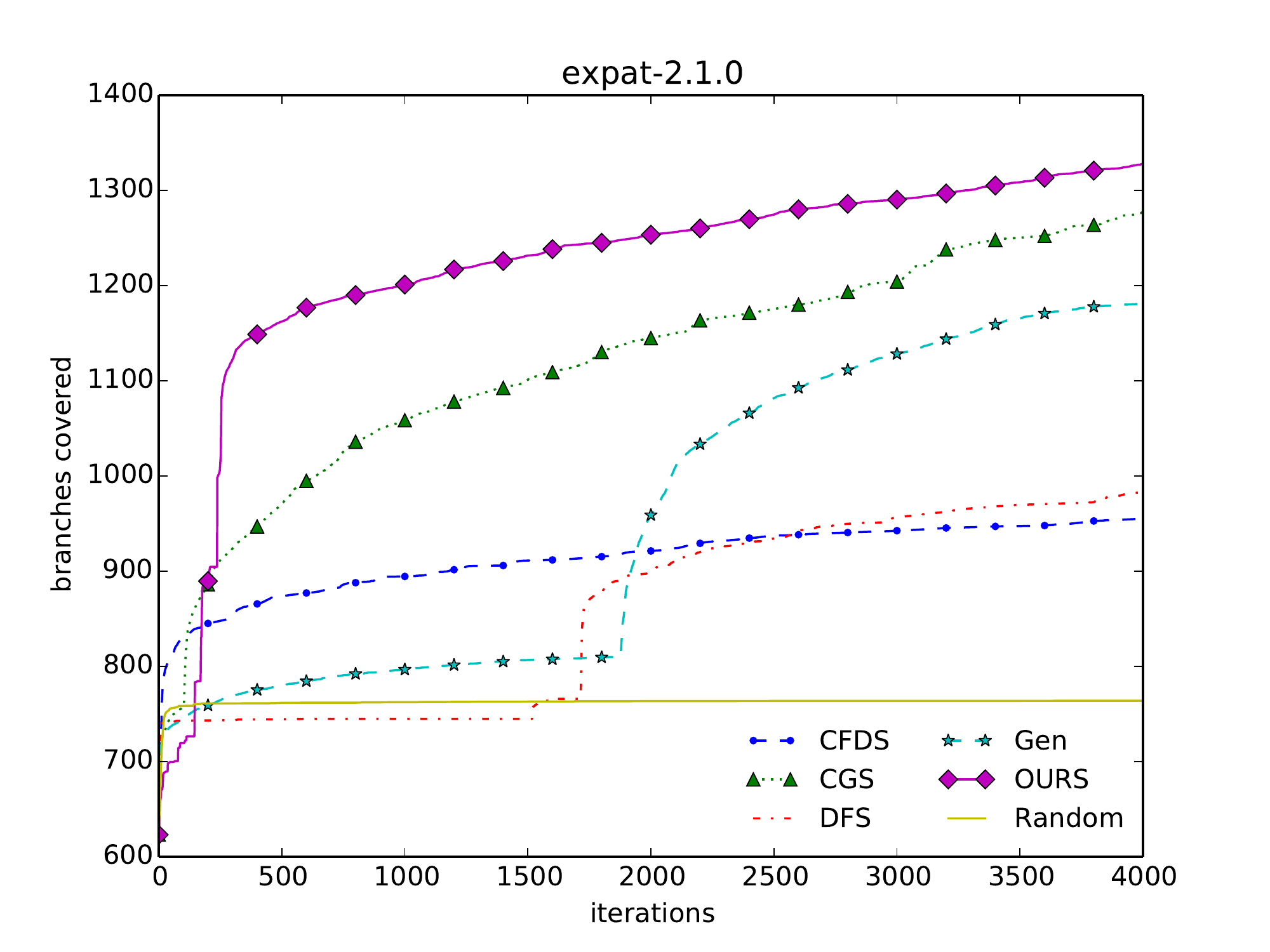} &
		\includegraphics[clip, trim={0.8cm 0.4cm 1.0cm 0.5cm}, width=8cm]{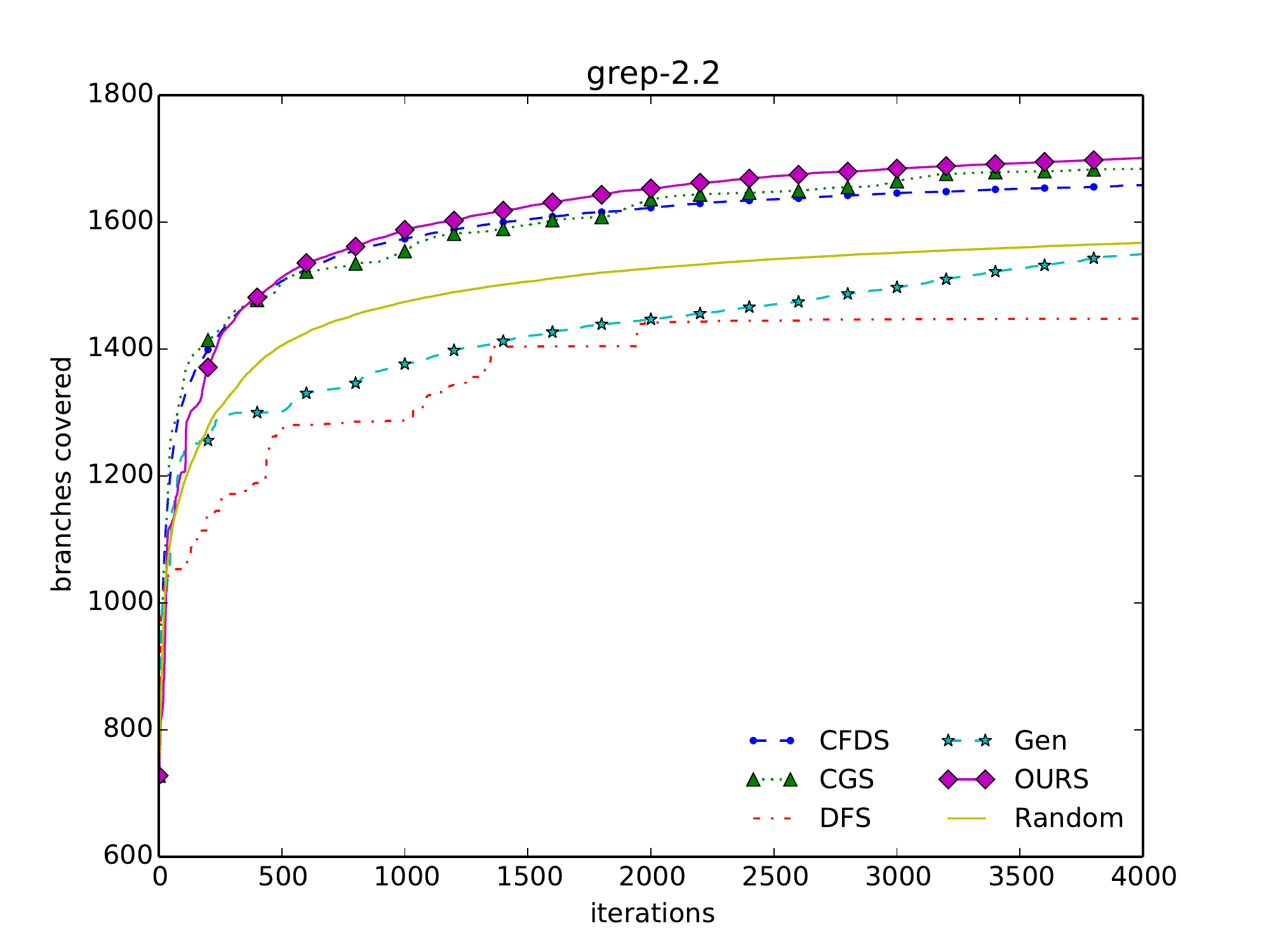}\\
		\includegraphics[clip, trim={0.8cm 0.4cm 1.0cm 0.5cm}, width=8cm]{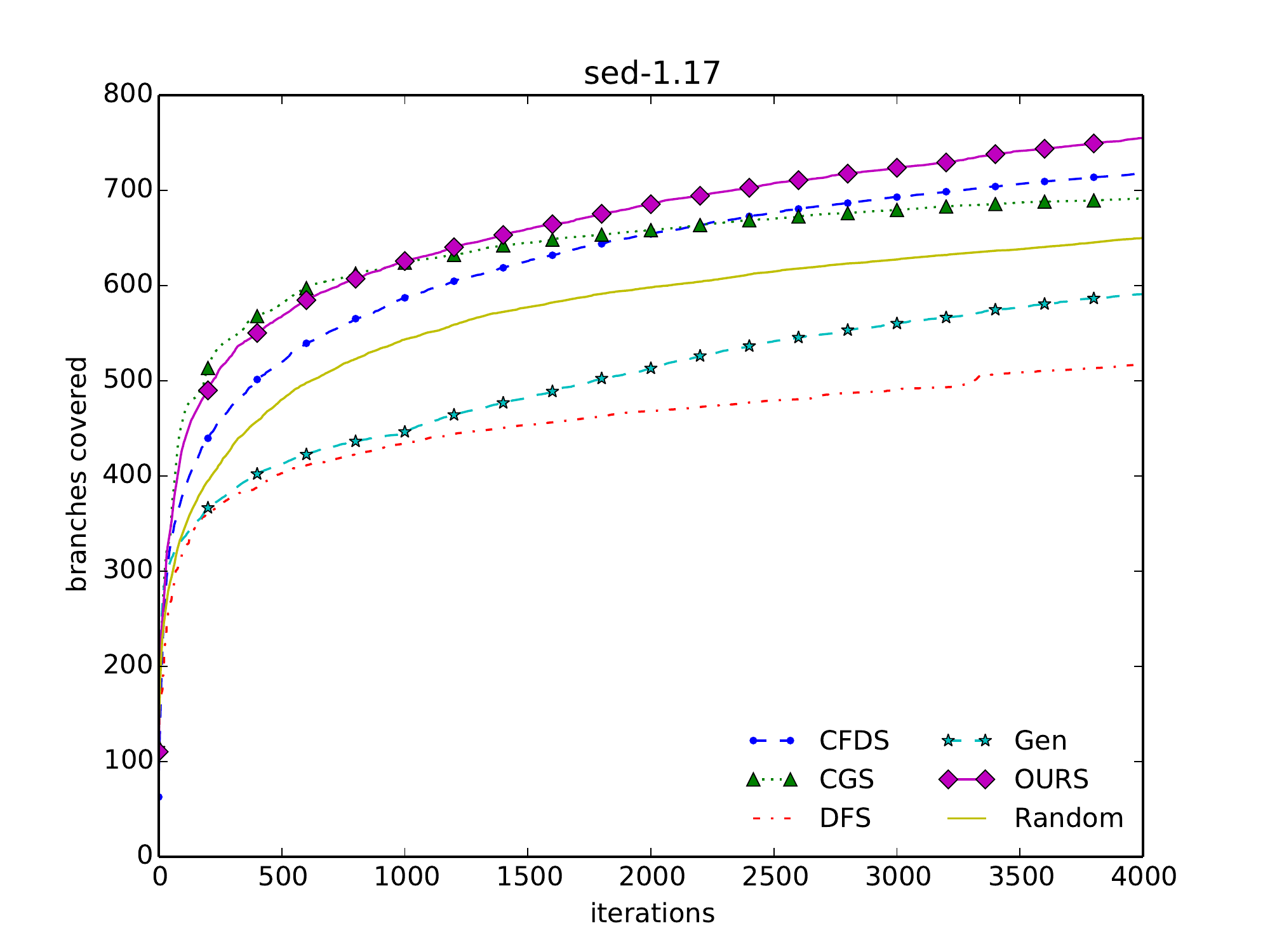} &
		\includegraphics[clip, trim={0.8cm 0.4cm 1.0cm 0.5cm}, width=8cm]{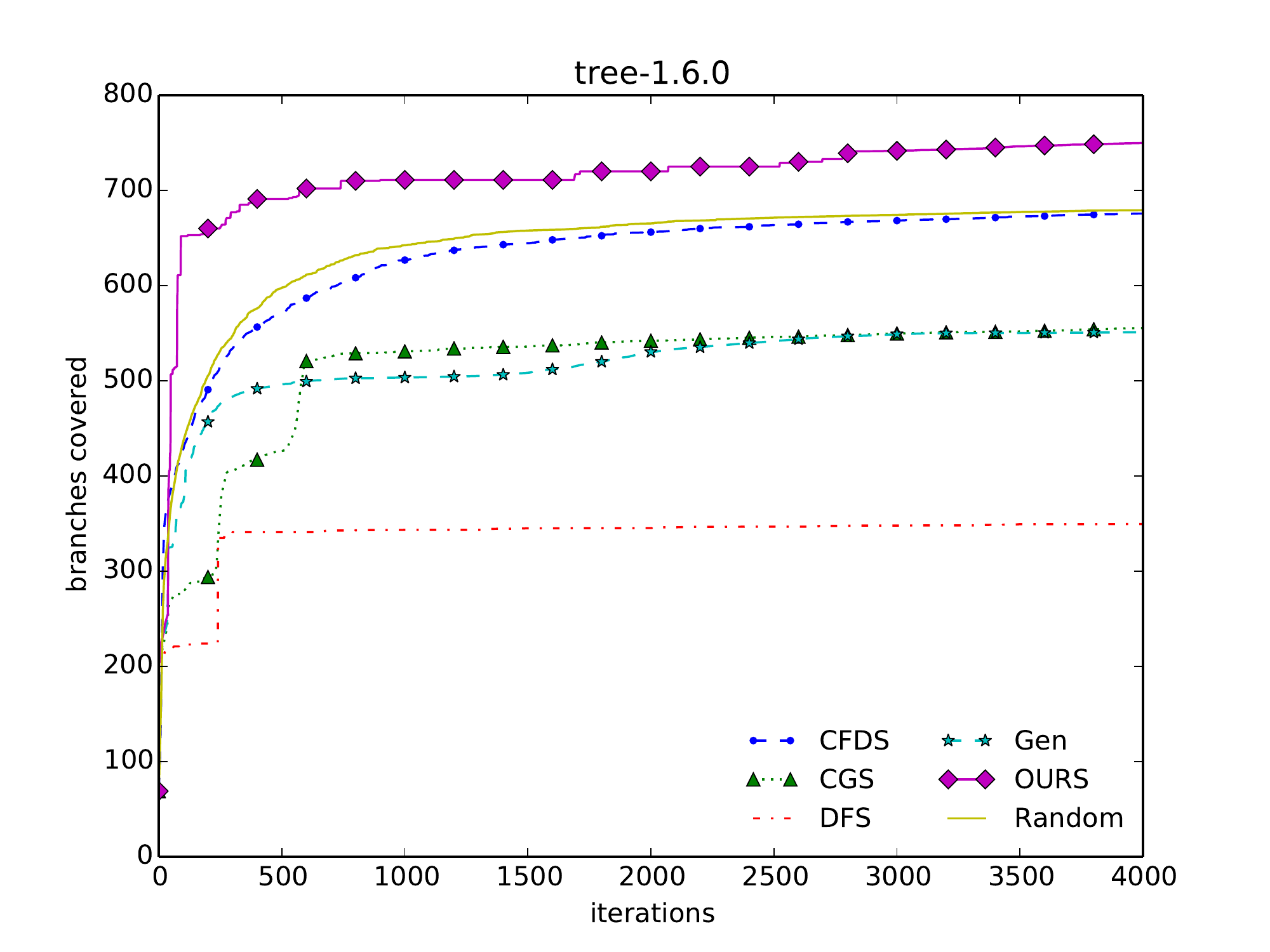}
	\end{tabular}
    \caption{Average branch coverage achieved by each search heuristic on 6 large benchmarks for concolic testing}
	\label{fig:AC}
\end{figure*}

\subsection{Effectiveness of Generated Heuristics}\label{sec:effectiveness}
For each benchmark program in Table~\ref{table:bench} and~\ref{table:kleebench}, we
ran our algorithm (Algorithm~\ref{alg:learning}) to generate our
search heuristic (ours), and compared its performance with that of the
existing heuristics in both approaches of dynamic symbolic execution.
We evaluate the effectiveness in terms of branch coverage.
For concolic testing, we also compare the heuristics in terms of bug
detection. 

\subsubsection{Concolic Testing}\label{sec:conbranch}

For branch coverage, we measured the average and maximum coverages.
The average branch coverage is obtained by averaging the results over
the 100 trials (50 for {\tt vim}).  The maximum coverage refers to the
highest coverage achieved during the 100 trials (50 for {\tt vim}).
The former indicates the average performance while the latter the best
performance achievable by each heuristic.

Figure~\ref{fig:AC} compares the average branch coverage achieved by
different search heuristics on 6 large benchmarks in Table~\ref{table:bench}.
The results show that the search heuristics generated by our approach (ours)
achieve the best coverage on all programs. In particular, ours significantly
increased the branch coverage on two largest benchmarks: \texttt{vim} and
\texttt{gawk}. For \texttt{vim}, ours covered 8,297 branches in 4,000
executions while the CFDS heuristic, which took the second place for {\tt
vim}, covered 7,990 branches.  Note that CFDS is already highly tuned and
therefore outperforms the other heuristics for {\tt vim} (for instance, CGS
covered 6,166 branches only).  For \texttt{gawk}, ours covered 2,684 branches
while the CGS heuristic, the second best one, managed to cover 2,321
branches. For {\tt expat}, {\tt sed}, and {\tt tree}, our approach improved
the existing heuristics considerably. For example, ours covered 1,327
branches for \texttt{expat}, increasing the branch coverage of CGS by 50. For
\texttt{grep}, ours also performed the best followed by CGS and CFDS. On
small benchmarks, we obtained similar results; ours (together with CGS)
consistently achieved the highest average coverage (Table~\ref{tb:small}). In
the rest of the paper, we focus only on the 6 large benchmarks, where
existing manually-crafted heuristics fail to perform well.

On all benchmarks in Figure~\ref{fig:AC}, ours exclusively covered branches that
were not covered by other heuristics. For example, in {\tt vim},
a total of 504 branches were exclusively covered by our heuristic.
For other programs, the numbers are: {\tt expat}(14), {\tt gawk}(7),
{\tt grep}(23), {\tt sed}(21), {\tt tree}(96).

These results are statistically significant: on all benchmark
programs in Figure~\ref{fig:AC}, the $p$ value was less than $0.01$
according to Wilcoxon signed-rank test.
In Figure~\ref{fig:AC}, the standard deviations for each heuristic
are as follows:
(1) {OURS}: {\tt vim}(258), {\tt expat}(42), {\tt gawk}(0), {\tt
  grep}(51), {\tt sed}(22), {\tt tree}(7);
(2) {CFDS}: {\tt vim}(252), {\tt expat}(44), {\tt gawk}(120), {\tt
  grep}(33), {\tt sed}(24), {\tt tree}(13);
(3) {CGS}: {\tt vim}(200), {\tt expat}(24), {\tt gawk}(57), {\tt
  grep}(29), {\tt sed}(27), {\tt tree}(15).
Other search heuristics also have similar standard deviations.

\begin{figure}[]
	\centering
	\begin{tabular}{cc}
		\includegraphics[clip, trim={0.8cm 0.4cm 1.0cm 0.5cm}, width=8cm]{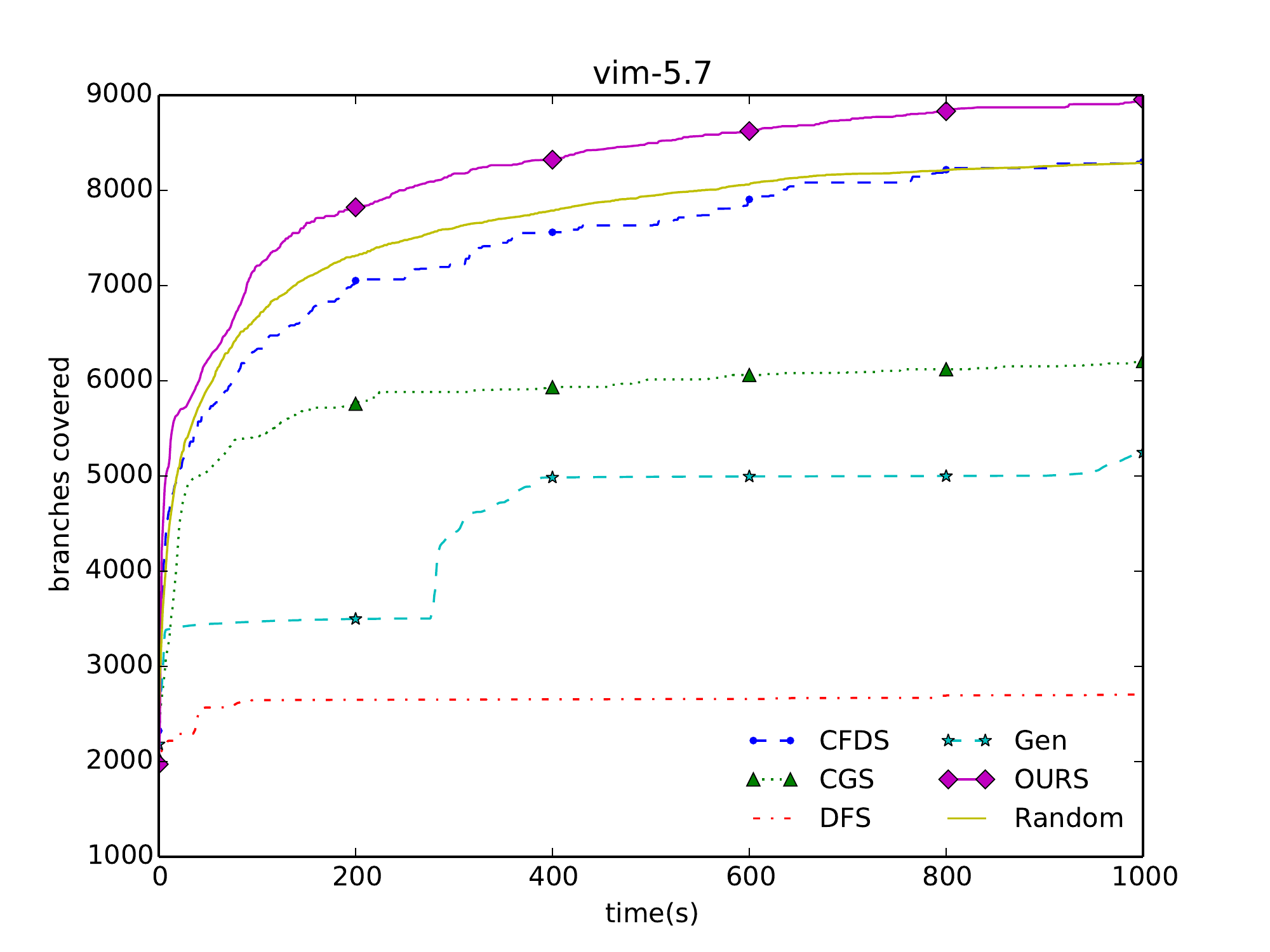}
	\end{tabular}
    \caption{Performance w.r.t. execution time (vim-5.7)}
	\label{fig:vim_time}
\end{figure}

In Figure~\ref{fig:AC}, we compared the effectiveness of search
heuristics over iterations (\# of executions),
but our approach was also superior to
others over execution time. For example, given the same time budget
(1,000 sec), ours and Random (the second best)
covered 8,947 and 8,272 branches, respectively, for {\tt vim}
(Figure~\ref{fig:vim_time}). The results were averaged over 50
trials.

Table~\ref{tb:best} compares the heuristics in terms of the maximum
branch coverage on 6 large benchmarks. The results show that our approach
in this case also achieves the best performance on all programs.
For instance, in \texttt{vim}, we considerably increased the coverage of CFDS,
the second best strategy; ours covered 8,788 branches while CFDS managed
to cover 8,585. For \texttt{expat}, ours and CGS (the second best) have
covered 1,422 and 1,337 branches, respectively.

Note that there is no clear winner among the existing search
heuristics.  Except for ours, CFDS took the first place for {\tt vim}
and {\tt sed} in terms of average branch coverage. For {\tt gawk},
{\tt expat}, and {\tt grep}, the CGS heuristic was the best. For {\tt
  tree}, the Random heuristic was better than CFDS and CGS.
In terms of the maximum branch coverage, CFDS was better than the
others for {\tt vim} and {\tt gawk} while CGS was for {\tt grep} and
{\tt sed}. The Gen and Random heuristics surpassed CFDS and CGS in {\tt
  expat} and {\tt tree}, respectively.

\begin{table}[]
\centering
\caption{Average branch coverage on 4 small benchmarks}
\label{tb:small}
\scalebox{0.9}{
\begin{tabular}{@{}crrrrrr@{}}
\toprule
\multicolumn{1}{l}{} & \multicolumn{1}{c}{\textbf{OURS}} & \multicolumn{1}{c}{CFDS} & \multicolumn{1}{c}{\textbf{CGS}} & \multicolumn{1}{c}{Random} & \multicolumn{1}{c}{Gen} & \multicolumn{1}{c}{DFS} \\ \midrule
cdaudio              & \textbf{250}                      & \textbf{250}             & \textbf{250}                     & 242                        & \textbf{250}            & 236                     \\
floppy               & \textbf{205}                      & \textbf{205}             & \textbf{205}                     & 170                        & \textbf{205}            & 168                     \\
replace              & \textbf{181}                      & 177                      & \textbf{181}                     & 174                        & 176                     & 171                     \\
kbfiltr              & \textbf{149}                      & \textbf{149}             & \textbf{149}                     & \textbf{149}               & \textbf{149}            & 134                     \\ \bottomrule
\end{tabular}}
\end{table}

\begin{table}[]
\centering
\caption{Effectiveness in terms of maximum coverage}
\label{tb:best}
\scalebox{0.9}{
\begin{tabular}{@{}crrrrrr@{}}
\toprule
      & \multicolumn{1}{c}{\textbf{OURS}} & \multicolumn{1}{c}{CFDS} & \multicolumn{1}{c}{CGS} & \multicolumn{1}{c}{Random} & \multicolumn{1}{c}{Gen} & \multicolumn{1}{c}{DFS} \\ \midrule
vim   & \textbf{8,788}                    & 8,585                    & 6,488                   & 8,143                      & 5,161                   & 2,646                   \\
expat & \textbf{1,422}                    & 1,060                    & 1,337                   & 965                        & 1,348                   & 1,027                   \\
gawk  & \textbf{2,684}                    & 2,532                    & 2,449                   & 2,035                      & 2,443                   & 1,025                   \\
grep  & \textbf{1,807}                    & 1,726                    & 1,751                   & 1,598                      & 1,640                   & 1,456                   \\
sed   & \textbf{830}                      & 780                      & 781                     & 690                        & 698                     & 568                     \\
tree  & \textbf{797}                      & 702                      & 599                     & 704                        & 600                     & 360                     \\ \bottomrule
\end{tabular}}
\end{table}

\begin{figure*}[h]
	\centering
	\begin{tabular}{cc}
		\includegraphics[clip, trim={0.8cm 0.4cm 1.0cm 0.5cm}, width=8cm]{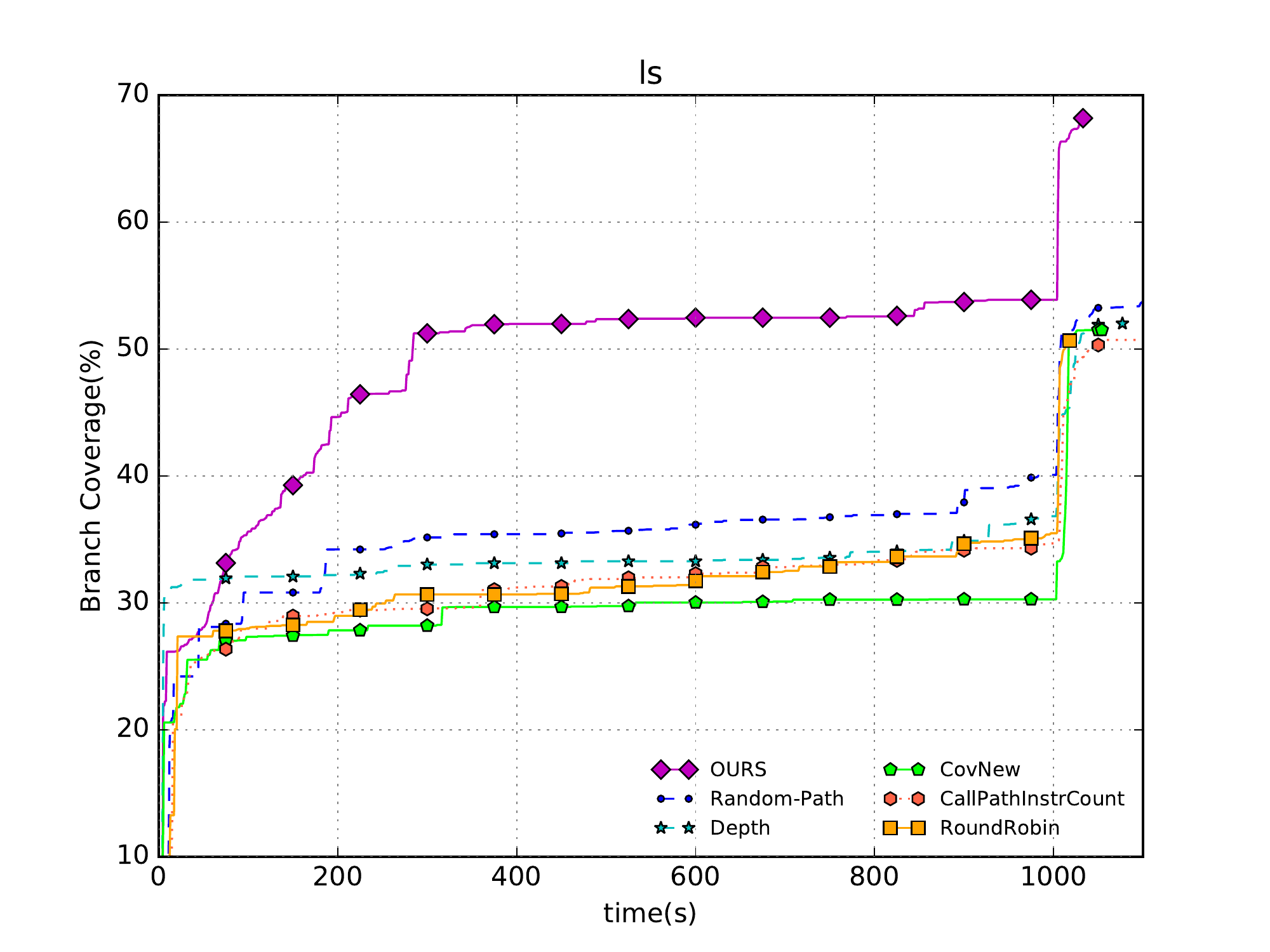} &
		\includegraphics[clip, trim={0.8cm 0.4cm 1.0cm 0.5cm}, width=8cm]{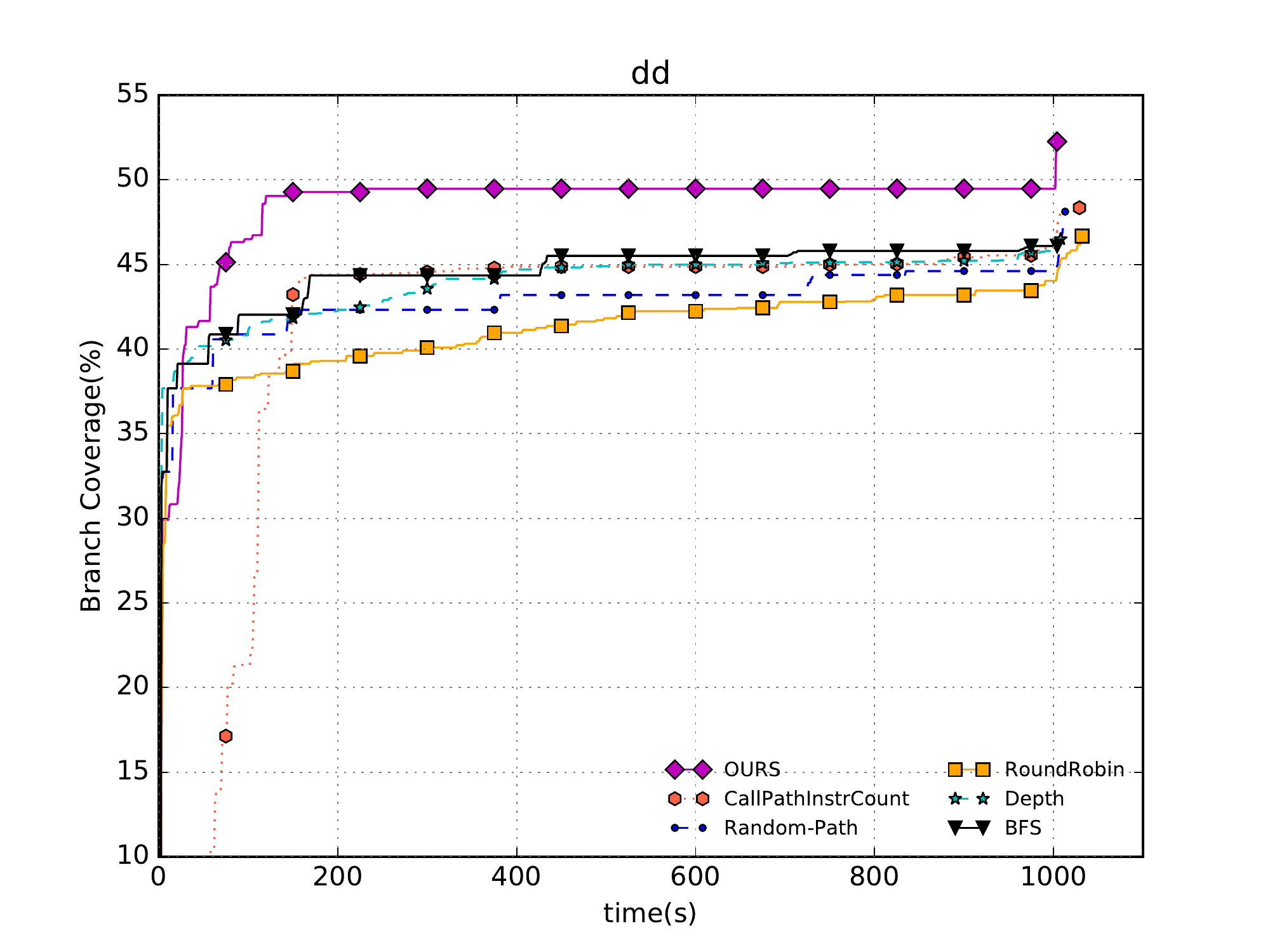} \\
		\includegraphics[clip, trim={0.8cm 0.4cm 1.0cm 0.5cm}, width=8cm]{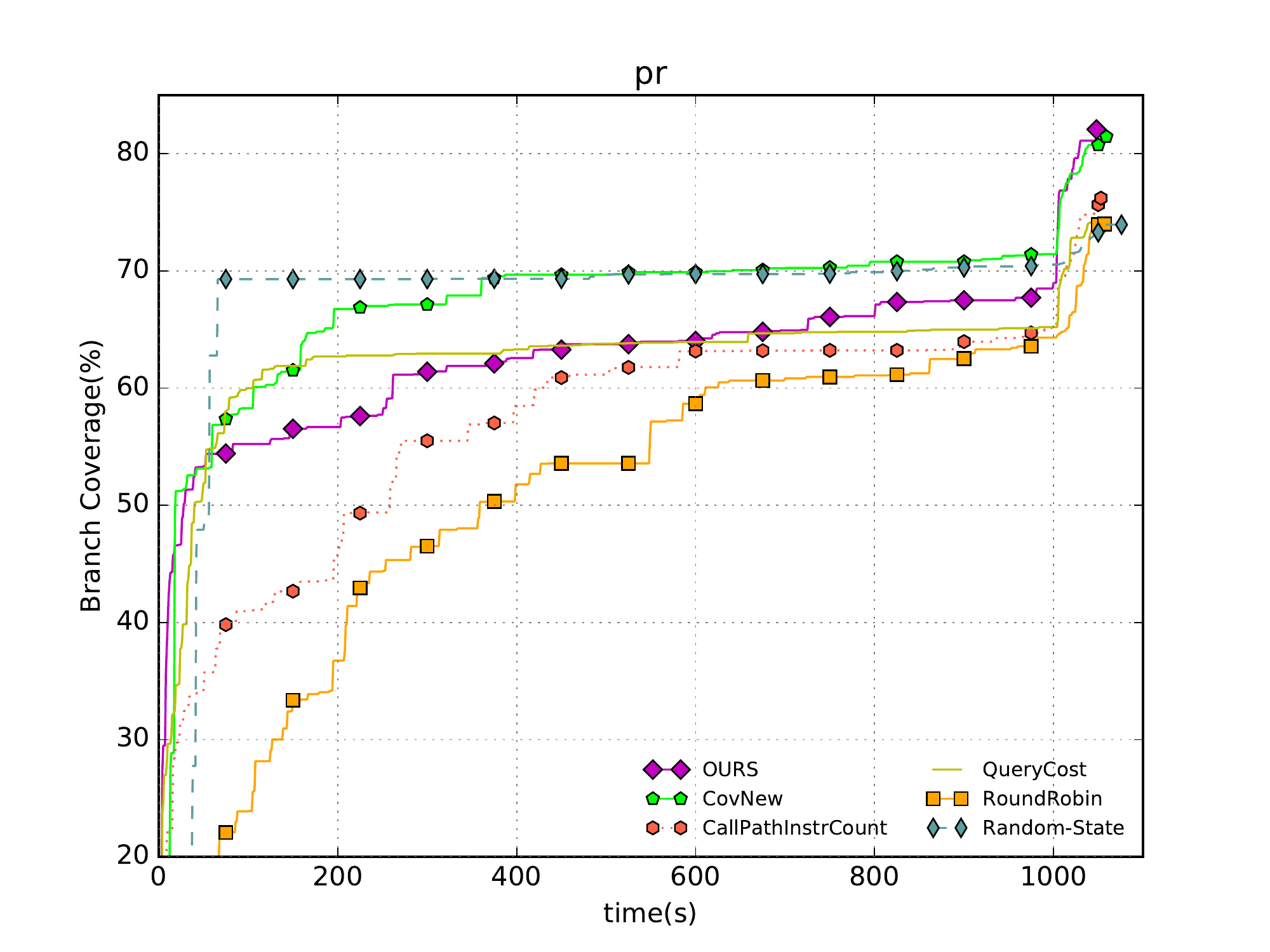} &
		\includegraphics[clip, trim={0.8cm 0.4cm 1.0cm 0.5cm}, width=8cm]{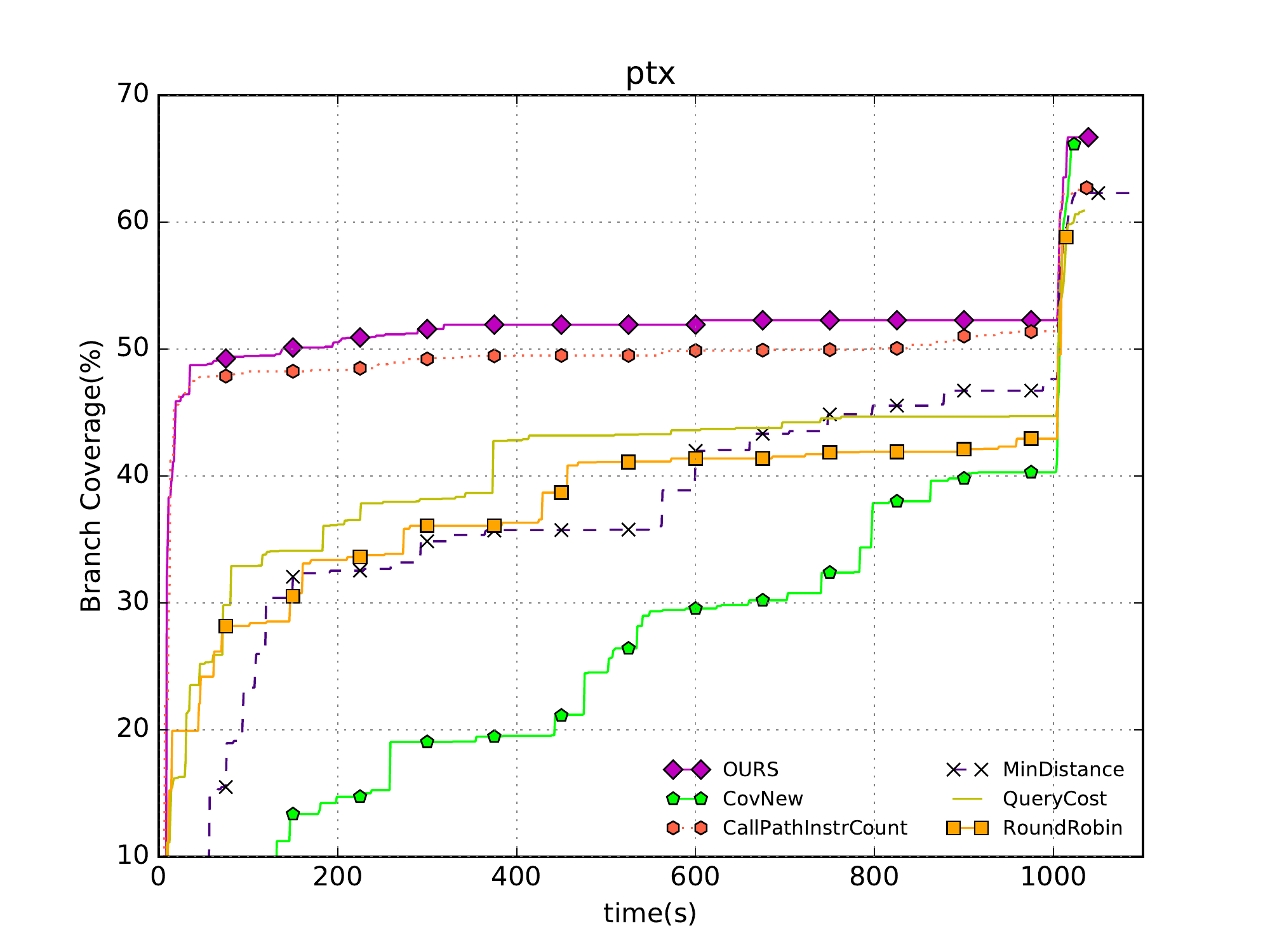}\\
		\includegraphics[clip, trim={0.8cm 0.4cm 1.0cm 0.5cm}, width=8cm]{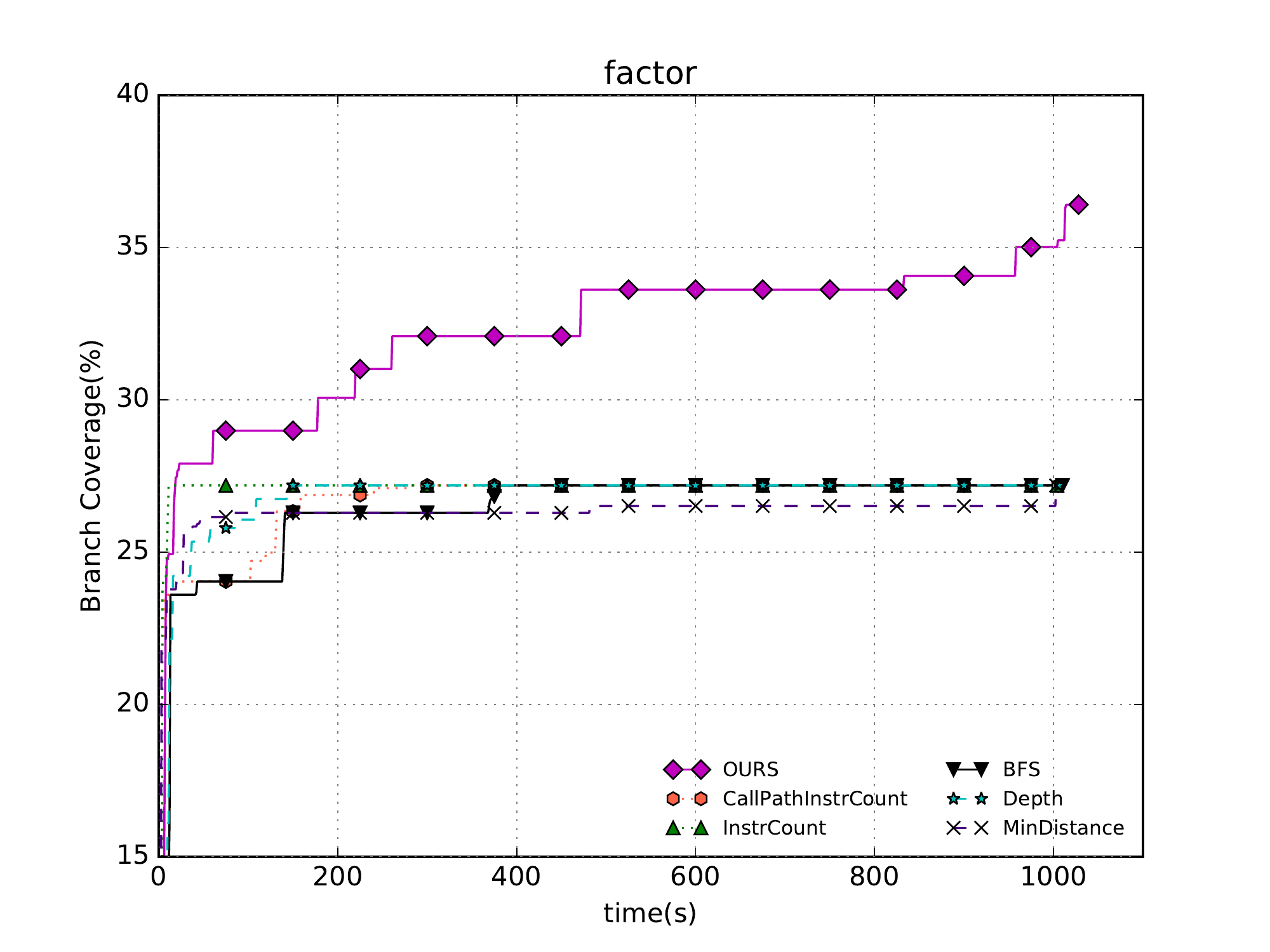} &
		\includegraphics[clip, trim={0.8cm 0.4cm 1.0cm 0.5cm}, width=8cm]{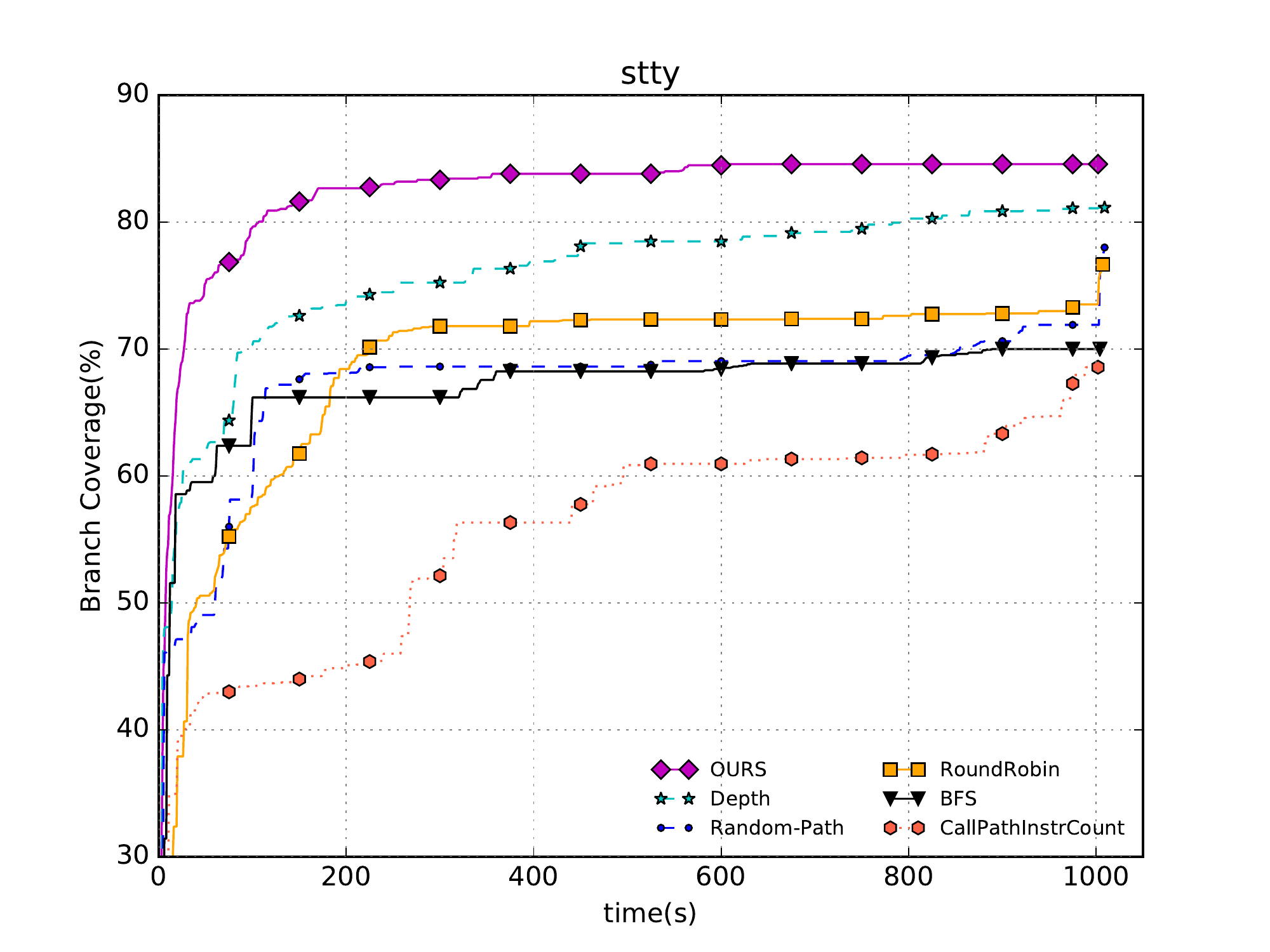}
	\end{tabular}
    \caption{Average branch coverage achieved by top 6 search heuristics on 6 large benchmarks for execution-generated testing}
	\label{fig:SymbolicAC}
\end{figure*}

\begin{table}[]
\centering
\caption{Effectiveness in terms of finding bugs}
\label{tb:bug}
\scalebox{0.9}{
\begin{tabular}{@{}ccccccc@{}}
\toprule
           & \textbf{OURS}   & CFDS     & CGS   & Random  & Gen    & DFS \\ \midrule
{\tt gawk-3.0.3} & \textbf{100/100} & 0/100   & 0/100 & 0/100  & 0/100  & 0/100   \\
{\tt grep-2.2}   & \textbf{47/100}  & 0/100   & 5/100 & 0/100  & 0/100  & 0/100   \\ \bottomrule
\end{tabular}}
\end{table}

We found that the increased branch coverage by our approach leads to
more effective finding of real bugs (not seeded ones).
Table~\ref{tb:bug} reports the number of trials
that successfully generate test-cases, which trigger
the known performance bugs in {\tt gawk} and {\tt grep}~\cite{gawkbug,grepbug}.
During the 100 trials (where a single trial consists of 4,000 executions), our heuristic always found the bug
in {\tt gawk} while all the other heuristics completely
failed to find it.
In {\tt grep}, ours succeeded to find the bug 47 times out of 100
trials, which is much better than CGS does (5 times). Other heuristics
were not able to trigger the bug at all.

Our heuristics are good at finding bugs because they are much better
than other heuristics in exercising diverse program paths.  We
observed that other heuristics such as CGS, CFDS, and Gen also covered the
branches where the bugs
originate. However, the bugs are caused only by some specific path
conditions and the existing heuristics could not generate
inputs that satisfy the conditions.

We remark that we did not specially tune our approach towards finding
those bugs.
In fact, we were not aware of the presence of those bugs
at the early stage of this work. The bugs in {\tt gawk} and {\tt grep}~\cite{gawkbug,grepbug} cause performance problems; for
example, {\tt grep-2.2} requires exponential time and memory on
particular input strings that involve
back-references~\cite{grepbug}. During concolic testing, we monitored
the program executions and restarted the testing procedure when the
subject program ran out of memory or time. Those bugs were
detected unexpectedly by a combination of this mechanism and our
search heuristic.


\subsubsection{Execution-Generated Testing}\label{sec:egtbranch}

$\paradyse$ also succeeded in generating the most effective search heuristic
for each program in Table~\ref{table:kleebench}, compared to all 11 
search heuristics implemented in KLEE. Here, we calculated the branch coverage with
respect to running time as follows:
\begin{enumerate}
\item While running KLEE on a program,
we recorded the creation time of each test-case generated by
Algorithm~\ref{alg:egt}. This step produces the data
$D = \myset{(T_i, t_i)}_{i=1}^M$,
where $T_i$ is a test-case, $t_i$ is its creation
time ($t_j < t_k$ if $j < k$), and $M$ is the number of generated test-cases.

\item When the time budget expires, we re-ran the 
  original binary of the program with each test-case $T_i$ in
  order. We computed the accumulated branch coverage $c_i$ of $T_i$ including
  the branches covered by all preceding test-cases $T_j (j <
  i)$. We plotted the data $\myset{(t_i, c_i)}_{i=1}^M$ to depict the
  coverage graph. To measure the branch coverage, we used
  \texttt{gcov}, a well-known tool for analyzing code coverage.
\end{enumerate}

Figure~\ref{fig:SymbolicAC} shows the average
branch coverage over the 10 trials achieved by top 6 heuristics on the
6 largest benchmarks. In
particular, our automatically-generated heuristic (ours) significantly increased the
average branch coverage for the largest benchmark {\tt ls}; ours covered 68\%
of the total branches in {\tt ls} while the second best heuristic
(Random-Path) only covered 53\% of the total branches. That is, ours is able
to cover about 227 more branches than the second best one on average during
the same time period. For {\tt dd}, {\tt factor}, and {\tt stty}, ours also
outperformed the existing heuristics. For example, in the case of {\tt
  factor}, ours was able to break the
coverage around 27\% that all the other heuristics eventually
converged on. For {\tt dd} and {\tt stty}, ours succeeded in
increasing the branch coverage by 4\% and 3\%, compared to the second best
heuristic of each benchmark: $\CallPathInstrCount$ ({\tt dd}) and $\Depth$ ({\tt stty}).

One interesting point in execution-generated testing is that there is a
significant increase in branch coverage at the end of the testing. For example,
Figure~\ref{fig:SymbolicAC} shows that all the search heuristics, including
ours, suddenly increase the branch coverage on {\tt ls} and {\tt ptx} when
the testing budget (1,000s) expires. This is due to the test-cases generated
after the testing budget is over; more precisely,
it is caused by lines 16--17 of Algorithm~\ref{alg:egt}.
The results indicate that the remaining states that have not yet reached the
end of the subject program (e.g., halt instruction in Algorithm~\ref{alg:egt})
contribute significantly to increasing branch coverage.

Likewise concolic testing, there is no obvious winner among 11 search
heuristics for execution-generated testing. Except for ours, there are the
four distinct heuristics which achieve the highest branch coverage at least
in one of the 6 benchmarks; the $\CovNew$ heuristic succeeded in
achieving the highest branch coverage for {\tt pr} and {\tt ptx}. Meanwhile,
$\CallPathInstrCount$ and $\Depth$ took the first place for {\tt dd} and {\tt
stty}, respectively. More surprisingly, the best heuristic for {\tt ls} is
the $\RandomPath$ heuristic that almost randomly picks a state from the
candidate states. These results on the existing heuristics for dynamic symbolic execution 
demonstrate our claim that manually-crafted heuristics are likely to be
suboptimal and unstable. On the other hand, our approach, $\paradyse$, is able to
consistently produce
the best search heuristics in both approaches of dynamic symbolic execution.


%% file: sec_5-3.tex
\begin{figure*}[h]
  \centering
  \begin{tabular}{cc}
    \includegraphics[clip, trim={0.5cm 0.7cm 0.7cm 0.2cm}, width=8cm]{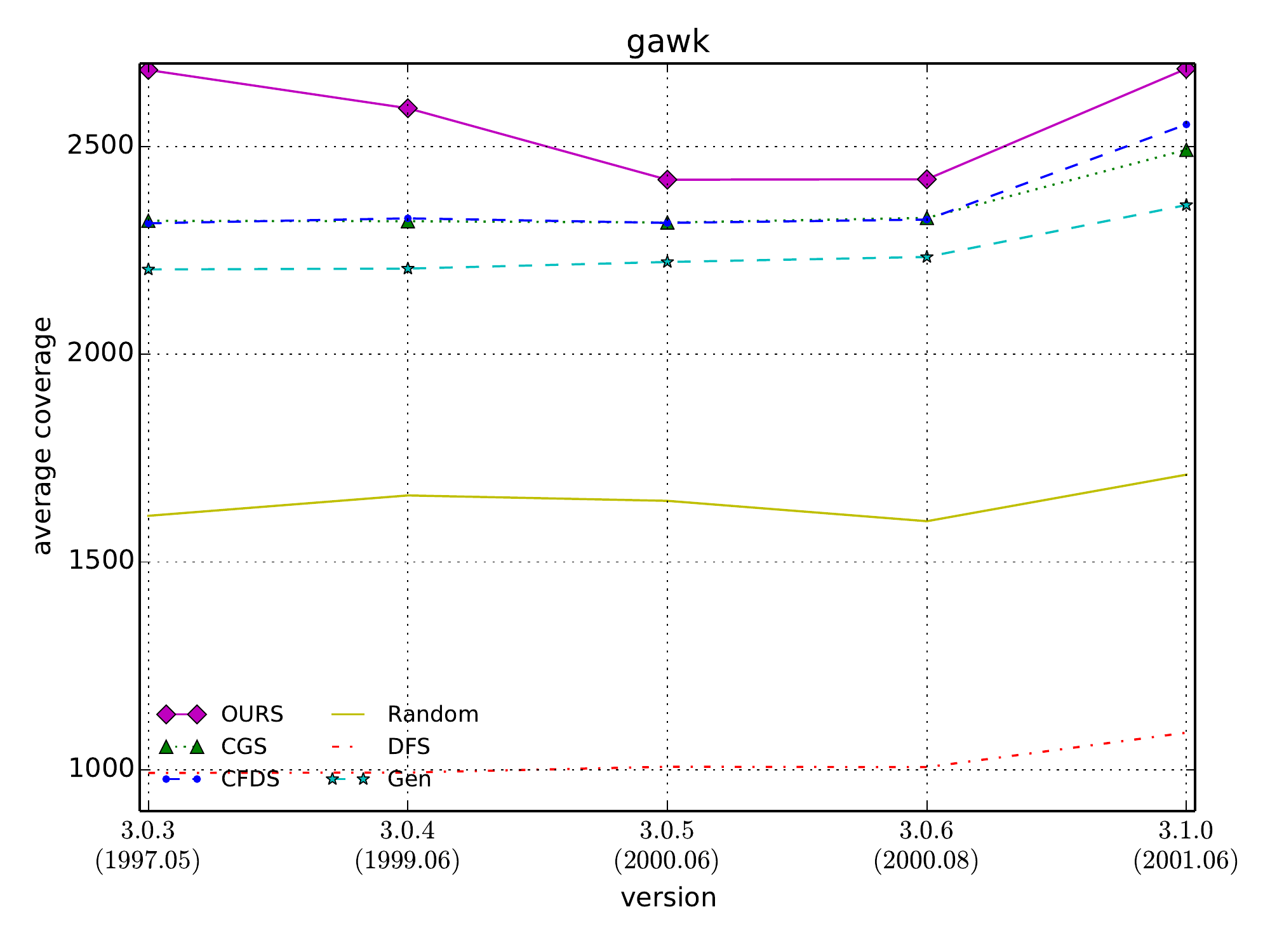} &
    \includegraphics[clip, trim={0.5cm 0.7cm 0.7cm 0.2cm}, width=8cm]{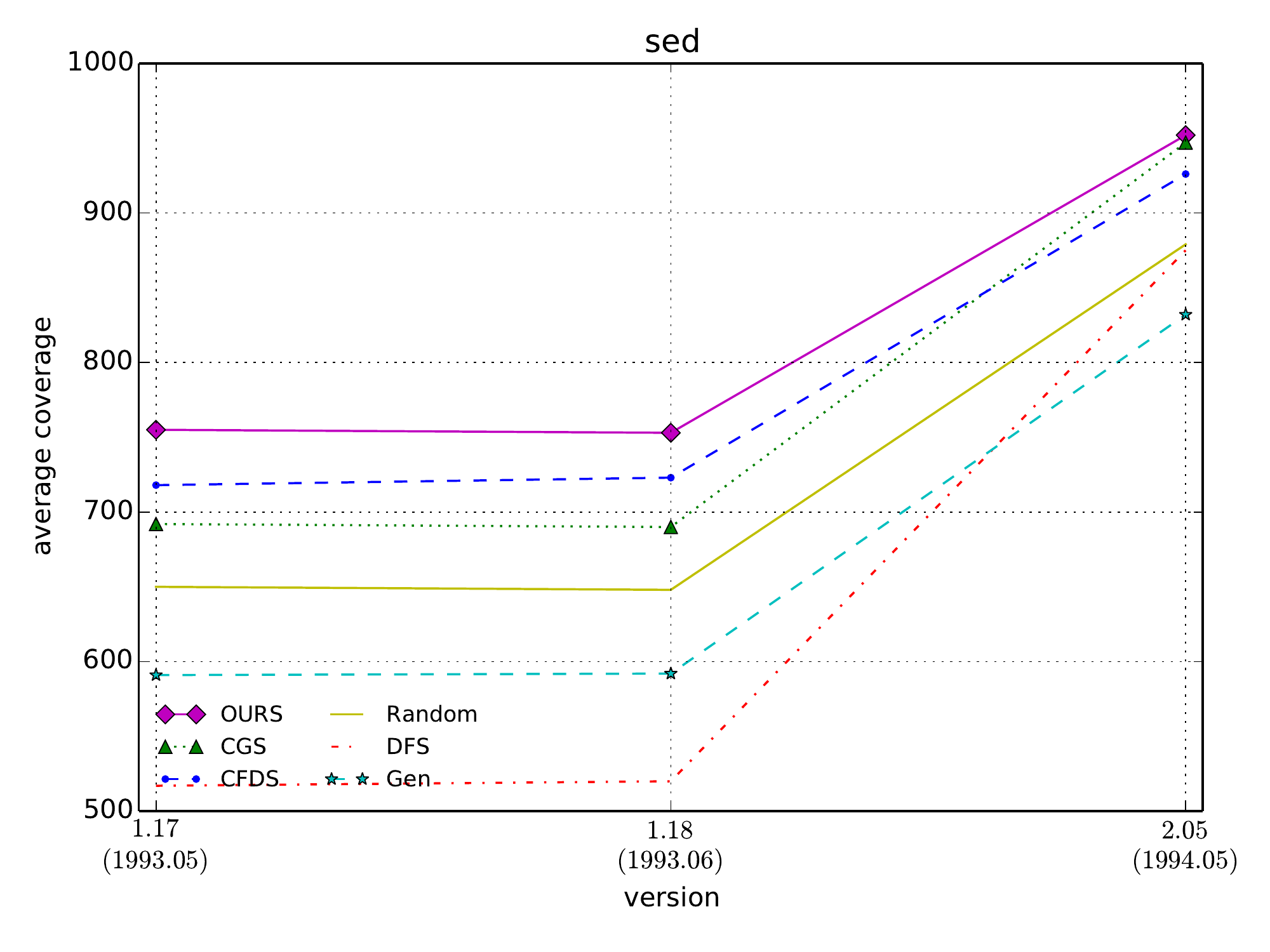}\\
  \end{tabular}
  \caption{Average coverage of each search heuristic for concolic testing on multiple
    subsequent program variants}
  \label{fig:evolve}
\end{figure*}

\subsection{Time for Obtaining the Heuristics}

Table~\ref{tb:time} reports the running time of our algorithm to generate the
search heuristics that were evaluated in Section~\ref{sec:effectiveness}. To obtain our
heuristics, we ran the optimization algorithm (Algorithm~\ref{alg:learning})
in parallel using 20 cores. For concolic testing, in the first
phase (`Find') of the algorithm, we sampled 1,000 parameters, where each core
is responsible for evaluating 50 parameters.  For {\tt vim}, we set the
sample size to 300 as executing {\tt vim} is expensive. For execution-generated testing,
we equally fixed the sample size to 200 because a single evaluation
(e.g, $\symbolic(P, \Choose_{\param_i}$)) is also expensive, where it took 1,000 seconds.
The results show that our algorithm converges within 2--6 iterations of the outer loop of
Algorithm~\ref{alg:learning}, taking 3--24 hours depending on the size
of the subject program and the number of iterations.

\begin{table}[t]
  \centering
  \caption{Time for generating the heuristics}
  \label{tb:time}
  \scalebox{0.9}{
  \begin{tabular}{@{}lrrr@{}}
    \toprule
    Benchmarks & \multicolumn{1}{l}{\# Sample} & \multicolumn{1}{l}{\# Iteration} & \multicolumn{1}{l}{Total times} \\ \midrule
    vim-5.7         & 300       & 5            & 24h 17min  \\
    expat-2.1.0     & 1,000     & 6            & 10h 25min  \\
    gawk-3.0.3      & 1,000     & 4            & 6h 28min   \\
    grep-2.2        & 1,000     & 5            & 5h 26min   \\
    sed-1.17        & 1,000     & 4            & 8h 55min   \\
    tree-1.6.0      & 1,000     & 4            & 3h 17min   \\ \midrule
    ls              & 200       & 4            & 17h 23min  \\
    dd              & 200       & 3            & 16h 52min   \\
    pr              & 200       & 5            & 21h 54min  \\
    ptx             & 200       & 2            & 11h 50min  \\
    factor          & 200       & 3            & 12h 43min  \\
    stty            & 200       & 4            & 16h 44min  \\ \bottomrule
  \end{tabular}}
\end{table}

\begin{table}[t]
  \renewcommand{\arraystretch}{1}%
  \centering
  \caption{Effectiveness in the training phase}
  \label{tb:acc}
  \scalebox{0.9}{
  \begin{tabular}{lrrrrrr}
    \toprule
              & \multicolumn{1}{c}{\textbf{OURS}} & \multicolumn{1}{c}{CFDS} & \multicolumn{1}{c}{CGS} & \multicolumn{1}{c}{Random} & \multicolumn{1}{c}{Gen} & \multicolumn{1}{c}{DFS} \\ \hline
    \multirow{1}{*}{\tt vim}         & \textbf{14,003}                   & 13,706                  & 7,934                   & 13,835                     & 7,290                   & 2,646                   \\
    \multirow{1}{*}{\tt expat}     & \textbf{2,455}                    & 2,339                   & 2,157                   & 1,325                      & 2,116                   & 2,036                   \\
    \multirow{1}{*}{\tt gawk}       & \textbf{3,473}                    & 3,382                   & 3,261                   & 3,367                      & 3,302                   & 1,905                   \\
    \multirow{1}{*}{\tt grep}        & \textbf{2,167}                    & 2,024                   & 2,016                   & 2,066                      & 1,965                   & 1,478                   \\
    \multirow{1}{*}{\tt sed}         & 1,019                             & 1,041                   & \textbf{1,042}          & 1,007                      & 979                     & 937                     \\
    \multirow{1}{*}{\tt tree}        & \textbf{808}                      & 800                     & 737                     & 796                        & 730                     & 665                     \\ \bottomrule
  \end{tabular}}
\end{table}

\begin{figure*}[h]
	\centering
	\begin{tabular}{c}
		\includegraphics[clip, trim={3.5cm 0.3cm 2.3cm 0.2cm}, width=15cm]{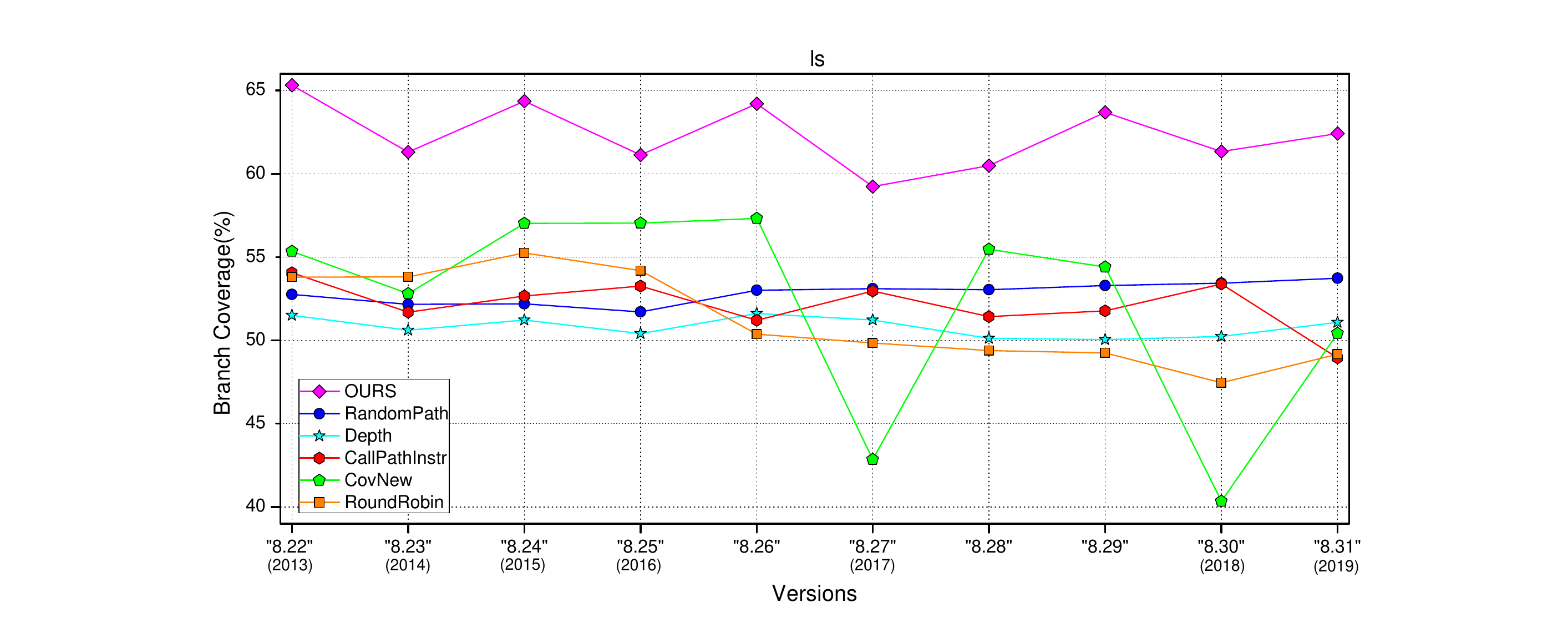}
	\end{tabular}
  \caption{Average coverage of each search heuristic for execution-generated testing on multiple subsequent program variants}
  \label{fig:evolve2}
\end{figure*}


Our approach requires training effort but it is rewarding because the learned
heuristic can be reused multiple times over a long period of time as the
subject program evolves. Moreover, we show that our approach enables effective
concolic testing even in the training phase.

\subsubsection{Reusability over Program Evolution}
The learned heuristics for both approaches of dynamic symbolic execution can
be reused over multiple subsequent program variations. To validate this
hypothesis in concolic testing, we trained a search heuristic on {\tt
gawk-3.0.3} and applied the learned heuristic to the subsequent versions
until {\tt gawk-3.1.0}. We also trained a heuristic on {\tt sed-1.17} and
applied it to later versions. Figure~\ref{fig:evolve} shows that the learned
heuristics manage to achieve the highest branch coverage
over the evolution of the programs. For example, ours covered at least 90
more branches than the second best heuristic (CFDS) in all variations between
{\tt gawk-3.0.3} and {\tt gawk-3.1.0}. The effectiveness lasted for at least
4 years for {\tt gawk} and 1 year for {\tt sed}.

For execution-generated testing, we also trained a search heuristic on the
largest benchmark {\tt ls} in GNU Coreutils-8.22 (2013) and applied it to the
subsequent, more precisely 9, versions from GNU Coreutils-8.23 (2014) to 8.31
(2019). Figure~\ref{fig:evolve2} shows that the learned heuristic succeeded
in achieving the highest branch coverage over all versions of GNU Coreutils.
The branch coverage difference between ours and the second best heuristic is
at least 4.1\% and up to 10.0\%. Note that ours consistently achieved the
highest coverage while the performance of the existing heuristics is
inconsistent with the evolution of the program; at the beginning of the
experiment, except for ours, we expected that the $\RandomPath$ heuristic
would be the best one, because $\RandomPath$ was the best in the recent
version of GNU Coreutils as we discussed in Section~\ref{sec:egtbranch}.
However, Figure~\ref{fig:evolve2} shows that except for ours, the $\CovNew$
heuristic is generally more effective than other search heuristics, including
$\RandomPath$, on multiple versions: 8.22, 8.24, 8.25, 8.26, 8.28 and 8.29.
Again, however, the performance of $\CovNew$ dropped sharply on the benchmark
{\tt ls} in GNU Coreutils-8.30; compared to the branch coverage achieved on
the immediate previous version (8.29), the branch coverage decreased by 14\%
in total. That is, existing search heuristics are likely to be
unstable across not only different programs but also different
versions of the same program.

\subsubsection{Effectiveness in the Training Phase}
Note that running Algorithm~\ref{alg:learning} is essentially running dynamic
symbolic execution on the subject program. Thus,
we compared the number of branches covered during this training phase in concolic
testing with the branches covered by other search heuristics given the same
time budget reported in Table~\ref{tb:time}.  Table~\ref{tb:acc} compares the
results: except for {\tt sed}, running Algorithm~\ref{alg:learning} achieves
greater branch coverage than others. To obtain the results for other
heuristics, we ran concolic testing (with $N=4,000$) repeatedly using the
same number of cores and amount of time.  For instance, in 24 hours,
Algorithm~\ref{alg:learning} covered 14,003 branches of {\tt vim} while
concolic testing with the CFDS and CGS heuristics covered 13,706 and 7,934 branches, respectively.



%% file: sec_5-4.tex
\begin{figure*}
        \centering
        \begin{subfigure}{\textwidth}
                \centering
                \includegraphics[clip, trim={0.2cm 0.1cm 0.2cm 0.2cm}, width=8cm]{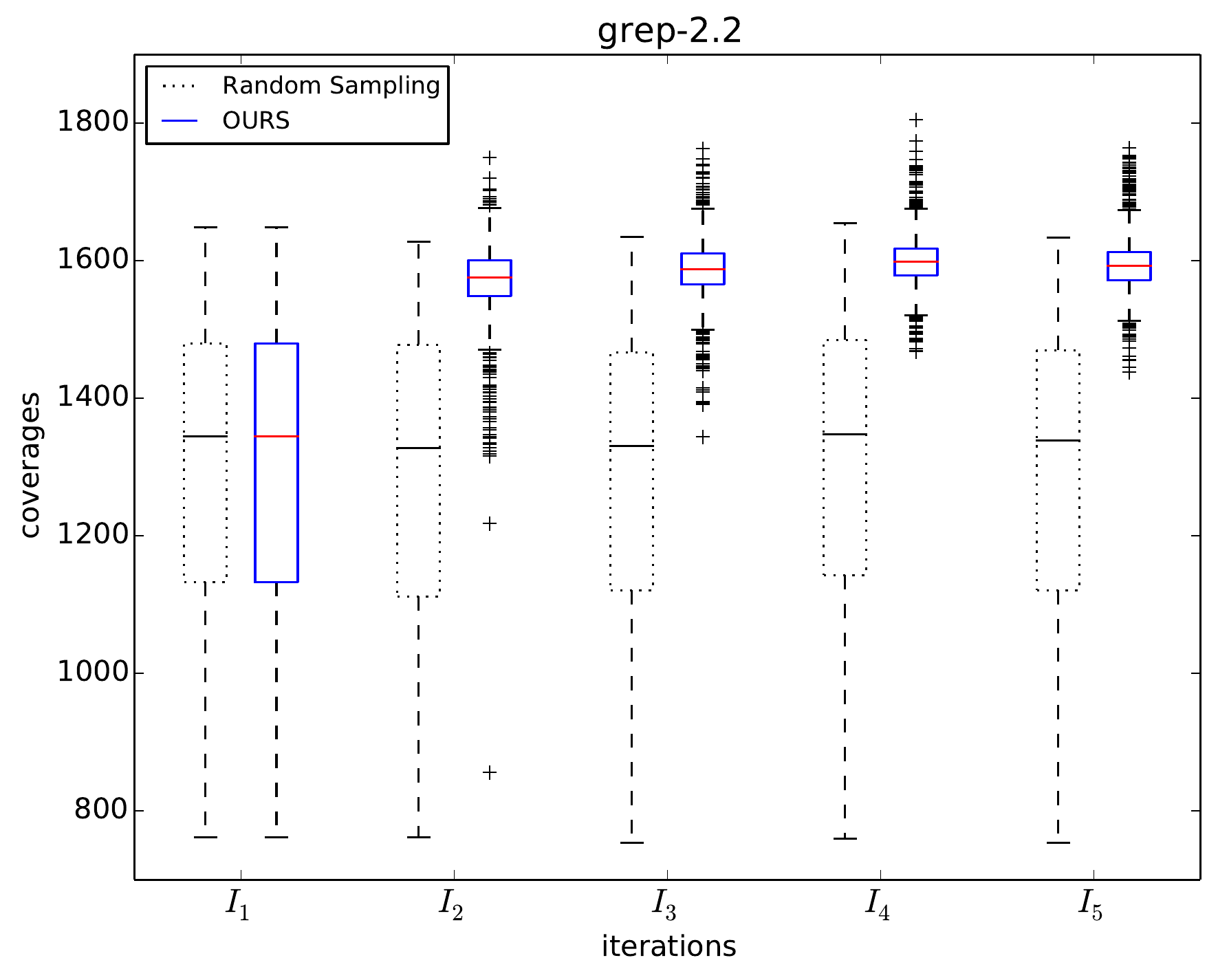}
                \includegraphics[clip, trim={0.2cm 0.1cm 0.2cm 0.2cm}, width=8cm]{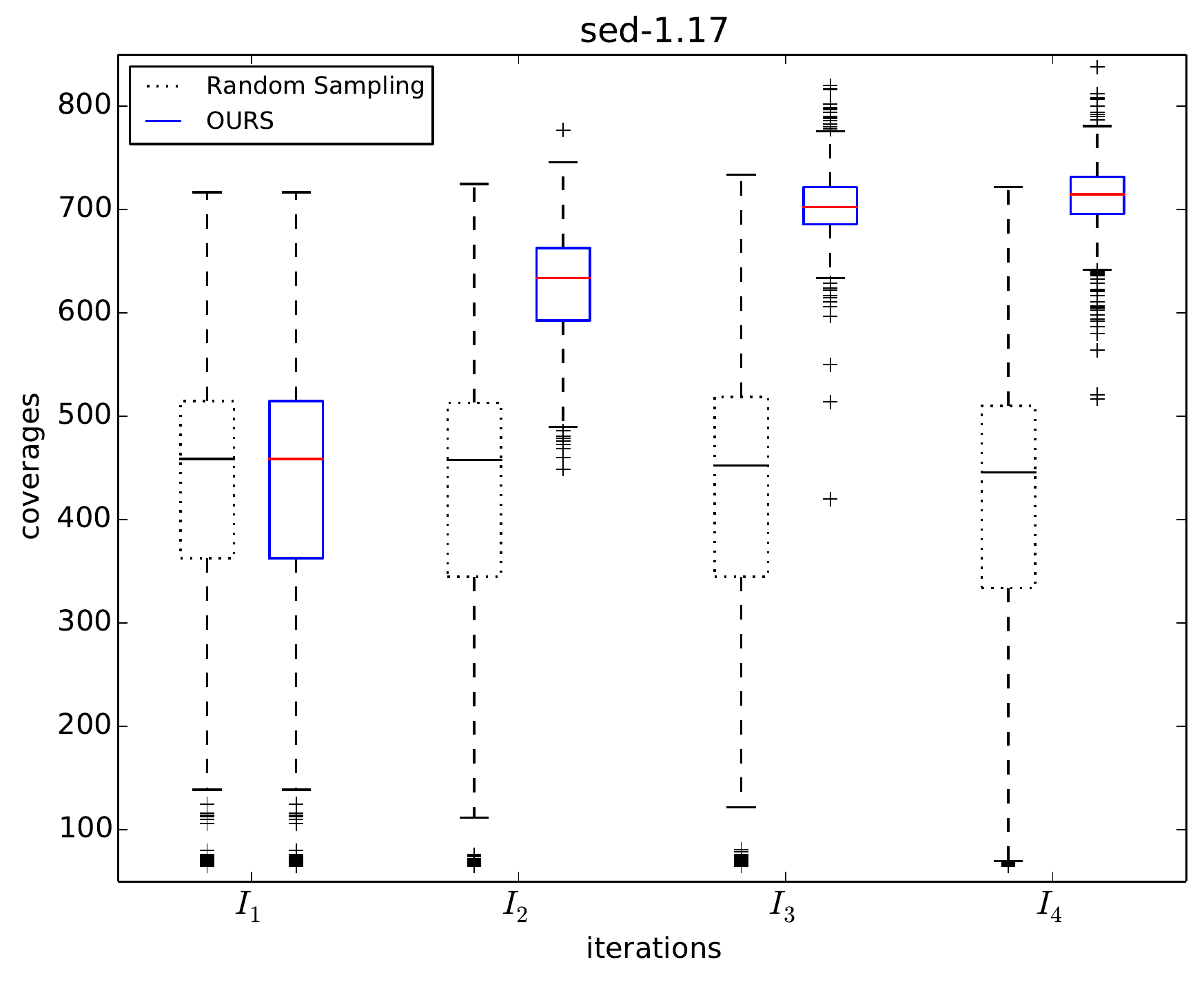}
                \caption{Concolic testing}
                \label{fig:concolicbox}
        \end{subfigure}
        \begin{subfigure}{\textwidth}
                \centering
                \includegraphics[clip, trim={0.2cm 0.1cm 0.2cm 0.2cm}, width=8cm]{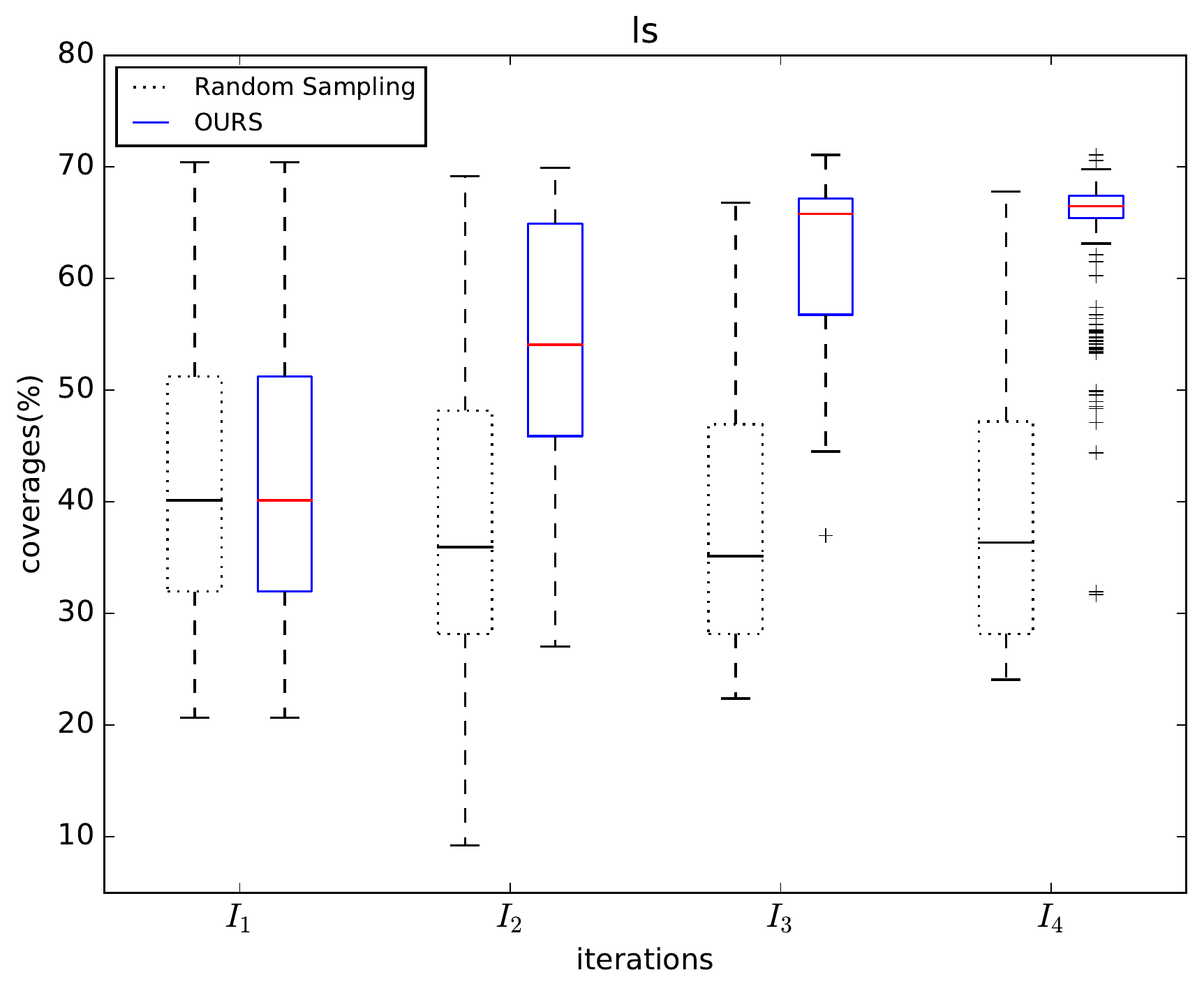}
                \includegraphics[clip, trim={0.2cm 0.1cm 0.2cm 0.2cm}, width=8cm]{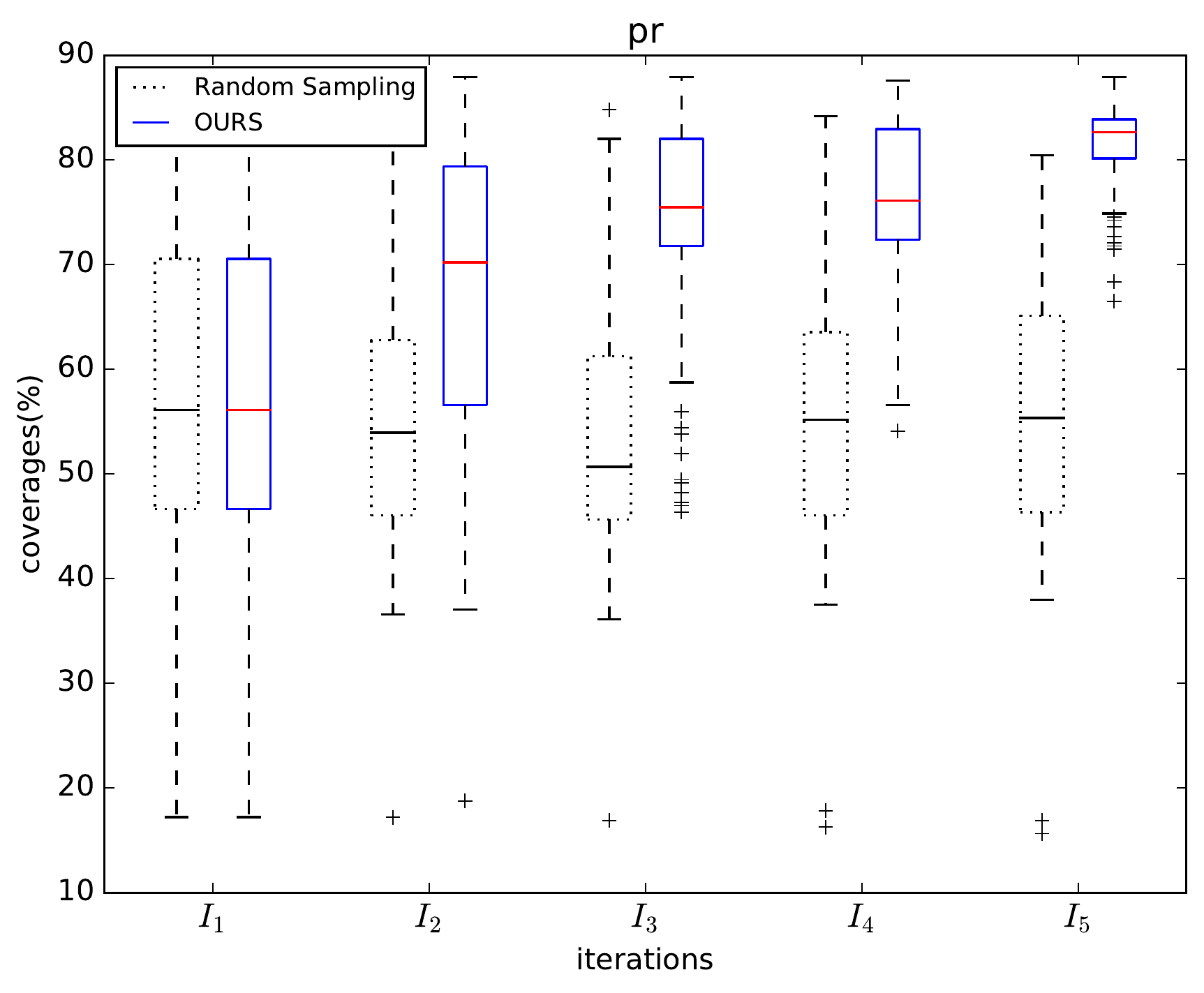}
                \caption{Execution-generated testing}
                \label{fig:egtbox}
        \end{subfigure}%
        \caption{Comparison of our learning algorithm and random sampling method}
        \label{fig:box}
\end{figure*}

\subsection{Efficacy of Optimization Algorithm}\label{sec:opt-algo}

We compared the performance of our optimization algorithm
(Algorithm~\ref{alg:learning}) with a naive approach based on random sampling
when generating search heuristics.
Because both approaches involve randomness, we statistically compare the
qualities of parameters found by our algorithm and the random sampling
method. We conducted the comparison on \texttt{grep-2.2} and
\texttt{sed-1.17} for concolic testing,
and \texttt{ls} and \texttt{pr} for execution-generated testing.

Figure~\ref{fig:box} shows the distributions of final coverages achieved by
those two algorithms instantiated in concolic testing and execution-generated
testing, respectively. First, for the former, our algorithm required a total
of 1,100 trials of concolic testing to complete a single refinement task: 100
trials for the Check phase to select top 2 parameters and the rest for the
Find phase to evaluate the parameters generated from the refined space.  We
compared the distributions throughout each iteration ($I_1, I_2, ..., I_N$)
where 1,100 trials were given as budget for finding parameters. Second, for
the latter, our algorithm needed a total of 300 trials of execution-generated
testing as a single refinement task: 200 trials (the Find phase) and 100
trials (the Check phase). That is, we compared the distributions for each
iteration with 300 trials. In both approaches of dynamic symbolic execution,
the first refinement task of our algorithm begins with the initial samples in
the first iteration $I_1$, which are prepared by random sampling method.

Figure~\ref{fig:concolicbox} and~\ref{fig:egtbox} show
that our algorithm is much superior to random sampling
method for both approaches of dynamic symbolic execution:
(1) the median of the samples increases while (2) the variance
decreases, as the refinement task in our algorithm goes on.  The
median value (the band inside a box) of the samples found by our
algorithm increases as the refinement task continues, while random
sampling has no noticeable changes. The increase of median indicates
that the probability to find a good parameter grows as the tasks
repeat. In addition, the variance (the height of the box, in simple)
in our algorithm decreases gradually, which implies that the mix of
{\it Check} and {\it Refine} tasks was effective.

We remark that use of our optimization algorithm was critical; the
heuristics generated by random sampling failed to surpass the existing
heuristics.  For instance, for \texttt{grep}, our
algorithm (Algorithm~\ref{alg:learning}) succeeded in generating a
heuristic which covered 1,701 branches on average. However, the best
one by random sampling covered 1,600 branches only, lagging behind CGS
(the second best) by 83 branches.

\begin{table*}
\centering
\caption{Top 10 features for concolic testing}
\label{tb:concolicFT}
\begin{subtable}{.5\textwidth}
\centering
\caption{Top 10 positive features}
\label{tb:concolicPF}
\scalebox{0.9}{
\begin{tabular}{@{}cllllll@{}}
\toprule
\multirow{2}{*}{Rank} & \multicolumn{6}{c}{Benchmarks}                                                                                                                                 \\ \cmidrule(l){2-7}
                      & \multicolumn{1}{c}{\tt vim} &
                                                      \multicolumn{1}{c}{\tt
                                                      gawk} &
                                                              \multicolumn{1}{c}{\tt
                                                              expat} &
                                                                       \multicolumn{1}{c}{\tt
                                                                       grep}
                                                                     &
                                                                       \multicolumn{1}{c}{\tt
                                                                       sed}
                                                                     &
                                                                       \multicolumn{1}{c}{\tt tree} \\ \midrule
1                     & \# 15                   & \textbf{\# 10($\star$)}  & \# 27                     & \# 14                    & \textbf{\# 13(+)}       & \# 36                    \\
2                     & \# 18                   & \textbf{\# 13(+)}        & \textbf{\# 30(+)}         & \# 40                    & \# 2                    & \# 15                    \\
3                     & \textbf{\# 35($\star$)} & \# 12                    & \# 23                     & \# 24                    & \# 29                   & \# 5                     \\
4                     & \# 40                   & \textbf{\# 38($\star$)}  & \textbf{\# 31(+)}         & \# 1                     & \# 3                    & \textbf{\# 25($\star$)}        \\
5                     & \textbf{\# 31(+)}       & \# 14                    & \# 4                      & \textbf{\# 30(+)}        & \# 8                    & \# 40                    \\
6                     & \# 7                    & \# 9                     & \# 9                      & \textbf{\# 38($\star$)}  & \textbf{\# 30(+)}       & \# 9                     \\
7                     & \textbf{\# 13(+)}       & \textbf{\# 35($\star$)}  & \# 8                      & \# 32                    & \textbf{\# 35($\star$)} & \textbf{\# 13(+)}        \\
8                     & \# 3                    & \textbf{\# 31(+)}        & \# 15                     & \# 17                    & \# 6                    & \# 39                    \\
9                     & \# 12                   & \# 4                     & \textbf{\# 25($\star$)}   & \textbf{\# 31(+)}        & \# 21                   & \textbf{\# 30(+)}        \\
10                    & \textbf{\# 10($\star$)} & \# 33                    & \# 7                      & \# 29                    & \# 16                   & \# 22                    \\ \bottomrule
\end{tabular}}

\end{subtable}%
\begin{subtable}{.5\textwidth}

\centering
\caption{Top 10 negative features}
\label{tb:concolicNF}
\scalebox{0.9}{
\begin{tabular}{@{}cllllll@{}}
\toprule
\multirow{2}{*}{Rank}  & \multicolumn{6}{c}{Benchmarks}                                                                                                                                 \\ \cmidrule(l){2-7}
                       & \multicolumn{1}{c}{\tt vim} &
                                                       \multicolumn{1}{c}{\tt
                                                       gawk} &
                                                               \multicolumn{1}{c}{\tt
                                                               expat}
                                                             &
                                                               \multicolumn{1}{c}{\tt
                                                               grep} &
                                                                       \multicolumn{1}{c}{\tt
                                                                       sed}
                                                                     &
                                                                       \multicolumn{1}{c}{\tt tree} \\ \midrule
1                      & \# 17                   & \textbf{\# 26(-)}        & \# 39                     & \# 20                    & \textbf{\# 11(-)}       & \textbf{\# 10($\star$)}        \\
2                      & \textbf{\# 11(-)}       & \# 8                     & \textbf{\# 35($\star$)}   & \# 39                    & \# 32                   & \textbf{\# 35($\star$)}        \\
3                      & \# 34                   & \# 16                    & \# 33                     & \textbf{\# 22(-)}        & \# 19                   & \# 6                     \\
4                      & \# 33                   & \# 29                    & \# 37                     & \textbf{\# 25($\star$)}  & \# 40                   & \# 24                    \\
5                      & \textbf{\# 22(-)}       & \# 3                     & \textbf{\# 38($\star$)}   & \textbf{\# 26(-)}        & \textbf{\# 38($\star$)} & \# 7                     \\
6                      & \# 21                   & \# 6                     & \# 2                      & \# 19                    & \# 18                   & \# 12                    \\
7                      & \textbf{\# 26(-)}       & \textbf{\# 22(-)}        & \# 24                     & \# 27                    & \# 5                    & \# 23                    \\
8                      & \textbf{\# 25($\star$)} & \textbf{\# 11(-)}        & \textbf{\# 22(-)}         & \# 21                    & \# 20                   & \# 2                     \\
9                      & \# 37                   & \# 19                    & \textbf{\# 10($\star$)}   & \# 33                    & \# 34                   & \# 27                    \\
\multicolumn{1}{l}{10} & \# 20                   & \# 28                    & \# 32                     & \# 37                    & \textbf{\# 26(-)}       & \textbf{\# 11(-)}        \\ \bottomrule
\end{tabular}}

\end{subtable}
\end{table*}


%% file: sec_5-5.tex
\begin{table*}
\centering
\caption{Top 5 features for execution-generated testing}
\label{tb:egtFT}
\begin{subtable}{.5\textwidth}
\centering
\caption{Top 5 positive features}
\label{tb:egtPF}
\scalebox{0.9}{
\begin{tabular}{@{}cllllll@{}}
\toprule
\multirow{2}{*}{Rank} & \multicolumn{6}{c}{Benchmarks}                                                                                                                               \\ \cmidrule(l){2-7}
                      & \multicolumn{1}{c}{\tt ls}  & \multicolumn{1}{c}{\tt dd}   & \multicolumn{1}{c}{\tt pr}    & \multicolumn{1}{c}{\tt ptx}  & \multicolumn{1}{c}{\tt factor} & \multicolumn{1}{c}{\tt stty} \\ \midrule
1                     & \textbf{\# 10($\star$)}     & \# 1                         & \# 19                         & \# 3                         & \textbf{\# 8($\star$)}         & \textbf{\# 10($\star$)} \\
2                     & \textbf{\# 24($\star$)}     & \# 3                         & \textbf{\# 24($\star$)}       & \textbf{\# 26($\star$)}      & \# 13                          & \textbf{\# 8($\star$)}  \\
3                     & \# 7                        & \textbf{\# 10($\star$)}      & \# 15                         & \# 1                         & \# 14                          & \textbf{\# 26($\star$)}  \\
4                     & \# 17                       & \textbf{\# 8($\star$)}       & \textbf{\# 26($\star$)}       & \textbf{\# 6($\star$)}       & \textbf{\# 4($\star$)}         & \textbf{\# 5($\star$)}   \\
5                     & \textbf{\# 4($\star$)}      & \# 25                        & \textbf{\# 23($\star$)}       & \textbf{\# 23($\star$)}      & \textbf{\# 6($\star$)}         & \textbf{\# 21($\star$)}  \\ \bottomrule
\end{tabular}}

\end{subtable}%
\begin{subtable}{.5\textwidth}
\centering
\centering
\caption{Top 5 negative features}
\label{tb:egtNF}
\scalebox{0.9}{
\begin{tabular}{@{}cllllll@{}}
\toprule
\multirow{2}{*}{Rank} & \multicolumn{6}{c}{Benchmarks}                                                                                                                               \\ \cmidrule(l){2-7}
                      & \multicolumn{1}{c}{\tt ls}  & \multicolumn{1}{c}{\tt dd}   & \multicolumn{1}{c}{\tt pr}    & \multicolumn{1}{c}{\tt ptx}  & \multicolumn{1}{c}{\tt factor} & \multicolumn{1}{c}{\tt stty} \\ \midrule
1                     & \textbf{\# 26($\star$)}     & \textbf{\# 4($\star$)}       & \textbf{\# 21($\star$)}       & \textbf{\# 10($\star$)}      & \# 19                          & \# 13                    \\
2                     & \# 20                       & \textbf{\# 5($\star$)}       & \textbf{\# 10($\star$)}       & \textbf{\# 21($\star$)}      & \textbf{\# 21($\star$)}        & \# 14                    \\
3                     & \# 18                       & \# 16                        & \# 12                         & \# 15                        & \textbf{\# 5($\star$)}         & \textbf{\# 24($\star$)}  \\
4                     & \textbf{\# 6($\star$)}      & \# 18                        & \# 22                         & \# 7                         & \textbf{\# 23($\star$)}        & \textbf{\# 4($\star$)}   \\
5                     & \textbf{\# 8($\star$)}      & \textbf{\# 21($\star$)}      & \# 20                         & \# 11                        & \# 22                          & \# 11                    \\ \bottomrule
\end{tabular}}
\end{subtable}
\end{table*}

\subsection{Important Features}

\subsubsection{Top-k Features}
We discuss the relative importance of features by analyzing the learned
parameters $\theta$ for each program in Table~\ref{table:bench}
and~\ref{table:kleebench}. Intuitively, when the $i$-th component $\theta^i$
has a negative number in $\theta$, it indicates that the branch having $i$-th
component should not be selected to be explored. Thus, both strong negative
and positive features are equally important for our approach to improve the
branch coverage. Table~\ref{tb:concolicFT} and Table~\ref{tb:egtFT} show the top-$k$
positive and negative features for concolic testing and execution-generated testing,
respectively; depending on the total number of features,
we set $k$ to 10 and 5 for the former and the latter.

The results show that there is no winning feature which always belongs to the
top-$k$ positive or negative features. Nevertheless, for concolic testing,
the features 13 (front parts of a path) and 30-31 (distances of uncovered
branches) are comparatively consistent positive ones. For 4 benchmarks, the
feature 11 (case statement), 22 (context) and 26 (frequently negated branch)
are included in the top 10 negative features. For designing effective search
heuristics, the key ideas of CFDS heuristic (\#30-31) and CGS (\#19-20, \#22)
heuristics are generally used as good positive and negative features,
respectively. {Surprisingly, in execution-generated testing,
there are no positive or negative features that are consistent in at least three benchmark programs.}

Many features for both approaches of dynamic symbolic execution simultaneously appear in
both positive and negative feature tables. That is, depending on the program
under test, the role of each feature changes from positive to negative (or
vice versa). In concolic testing, the features 10, 25, 35 and 38 appear in
both Table~\ref{tb:concolicPF} and Table~\ref{tb:concolicNF}. In particular,
the feature 10 is used as the most positive feature in \texttt{gawk} while it
is the most negative one for \texttt{tree}. In execution-generated testing,
the phenomenon is more prevalent;
the features 4, 5, 6, 8, 10, 21, 23, 24, and 26 serve as both positive and negative ones.
For instance, the feature 4 (branch inside a loop body) is
in top-5 positive one on {\tt ls} and {\tt factor}
while it is also in top-5 negative one on {\tt dd} and {\tt stty}.
This finding supports our claim that no single search heuristic can perform well for all benchmarks,
and therefore it should be tuned for each target program.

\subsubsection{Impact of Combining Static and Dynamic Features}

The combined use of static and dynamic features was important.
In concolic testing, we assessed the performance of our approach with different feature
sets in two ways: 1) with static features only; and 2) with dynamic
features only. Without dynamic features, generating good heuristic was feasible only for {\tt grep}.
Without static features, our approach succeeded in generating good heuristics for
{\tt grep} and {\tt tree} but failed to do so for the remaining
programs.


\subsection{Threats to Validity}
\begin{enumerate}
  \item {\bf Benchmarks}: For concolic testing, we collected eight benchmarks
  from prior work~\cite{Boonstoppel2008,Cadar2008, Burnim2008,Kim2011,
  Seo2014} and created two new benchmarks (\texttt{gawk} and \texttt{tree}).
  For execution-generating testing, we used the largest 6 benchmark programs
  in the latest version of GNU Coreutils, where it is the representative benchmark in prior
  work~\cite{Marinescu2012, Kuznetsov2012, Li2013, Yi2015, Yi2018, Wong2015,
  Mechtaev2018fse}. However, these 16 benchmarks may not be representative
  and not enough to evaluate the performance of the search heuristics in general.

  \item {\bf Testing budget}: For concolic testing, we chose 4,000 executions as the testing budget because it is
the same criterion that was used for evaluating the existing heuristics (CGS, CFDS) in prior work~\cite{Burnim2008,
  Seo2014}. For execution-generated testing, we set 1,000 seconds to the testing budget because
   using timing budget is common in previous works on KLEE~\cite{Cadar2008, Kim2012, Marinescu2012, Li2013}.
  However, this might not be the best setting in practice.

  \item {\bf Constraint solver}: The performance of dynamic symbolic execution may vary
depending on the choice of the SMT solver. For concolic testing, we used Yices~\cite{Dutertre06theyices}, the default SMT solver of CREST.
For execution-generated testing, we used STP~\cite{ganesh2007}, the default SMT solver of KLEE.
\end{enumerate}




%% file: relatedwork.tex
\section{Related work}

We discuss existing works on improving the performance of dynamic symbolic
execution. We classify existing techniques into the four classes: (1)
improving search heuristics; (2) hybrid approaches; (3) reducing
search space; (4) solving complex path conditions.
{Our work can also be seen as a combination of software testing and
  machine learning or search-based software testing.}


\subsection{Search Heuristics}
\label{sec:rel1}

As search heuristics are a critical component of dynamic symbolic execution, a
lot of techniques have been proposed.
However, all existing works on improving search heuristics focus on
manually-designing a new strategy~\cite{Seo2014,Burnim2008,Park2012,cabfuzz,
  Xie2009,Cadar2008, Li2013}.
In Section~\ref{sec:preliminaries}, we
already discussed the CFDS~\cite{Burnim2008} and CGS~\cite{Seo2014} heuristics.
Another successful heuristic is generational search~\cite{Godefroid2008}, which drives concolic testing
towards the highest incremental coverage gain to maximize code
coverage. For each execution path, all branches are negated and
executed. Then, next generation branch is selected according to the coverage gain of
each single execution. Xie et al.~\cite{Xie2009} designed a heuristic
that guides the search based on the fitness values that measure the
distance of branches in the execution path to the target branch. 
The CarFast heuristic~\cite{Park2012} guides concolic testing based on the number
of uncovered statements.
{The Subpath-Guided Search heuristic}~\cite{Li2013}{ steers
  symbolic execution to less explored areas of the subject program by using the length-$k$ subpath.}
Our work is different from these works as we automate
the heuristic-designing process itself.

\subsection{Hybrid Approaches}
\label{sec:rel2}
Our approach is orthogonal to the existing techniques that combine dynamic
symbolic execution with other testing techniques.
In~\cite{Majumdar2007,Garg2013}, techniques such as random testing are first
used and they switch to concolic testing when the performance gains saturate.
In~\cite{Inkumsah2008}, concolic testing is combined with evolutionary
testing to be effective for object-oriented programs.
{Munch}~\cite{Ognawala2018} {is a hybrid technique to combine
symbolic execution (e.g., KLEE}~\cite{Cadar2008}) {with fuzzing (e.g.,
AFL}~\cite{afl}) {to maximize the function coverage.}

\subsection{Reducing Search Space}
\label{sec:rel3}

Our work is also orthogonal to techniques that reduce the search space of symbolic execution~\cite{Boonstoppel2008,
  Jaffar2013,Godefroid2007,Godefroid2010,Daca2016,Trabish2018,Yi2015,Yi2018}.  
The read-write set
analysis~\cite{Boonstoppel2008} identifies and prunes program paths
that have the same side effects.
Jaffar et al.~\cite{Jaffar2013}
introduced an interpolation method that subsumes paths
guaranteed not to hit a bug. 
Goderfroid et al.~\cite{Godefroid2007,Godefroid2010} proposed to use
function summarizes to identify equivalence classes of function
inputs. It ensures that the concrete executions in the same class have
the same side effect.
Abstraction-driven concolic
testing \cite{Daca2016} also reduces search space for concolic testing
by using feedback from a model checker. 
ConTest~\cite{Cha2018template}{ aims to reduce the input space of concolic
testing by selectively maintaining symbolic variables via online learning.}
Chopper~\cite{Trabish2018}{ is a novel technique for performing
  symbolic execution
while safely ignoring functions of the subject program that users do not want to explore.}
Postconditioned symbolic execution~\cite{Yi2015,Yi2018}{  aims to prune redundant paths of the program
by using the post conditions accumulated during symbolic execution.}

{State-merging is a promising technique to reduce the number of states in
symbolic execution}~\cite{Zhang2018,Kuznetsov2012,Sen2015}.
Kuznetsov et al.~\cite{Kuznetsov2012}{ proposed a method to balance between reducing the
number of states and increasing the burden on the constraint solver by
statically and dynamically estimating the importance of the states.}
MultiSE~\cite{Sen2015}{ introduced a new technique to enable symbolic execution to merge states without
generating any auxiliary symbolic variables. Thereby, MultiSE is able to perform
symbolic execution even when it merges values that are unsupported by constraint solver in the states.}
Our work is orthogonal to state-merging techniques and can be combined with them to boost symbolic execution further.


\subsection{Solving Complex Path Conditions}
\label{sec:rel4}

Our technique can also be improved by incorporating existing
techniques for solving complex path conditions.
Conventional SMT solvers are not effective in handling constraints that involve non-linear arithmetic
or external function calls, which often causes symbolic execution to have poor
coverage.
In~\cite{Dinges2014}, an algorithm was introduced that can solve hard arithmetic constraints in path
conditions. The idea is to generate geometric structures that help
solve non-linear constraints with existing heuristics~\cite{Codognet2001}.
In \cite{Thome2017}, a technique to solve string constraints was
proposed based on ant colony optimization.
There are attempts to solve this problem by machine learning~\cite{Li2016}.
It encodes not only the simple linear path conditions, but also complex path
conditions (e.g., function calls of library methods) into the symbolic path
conditions. The objective function is defined by dissatisfaction degree. By
iteratively generating sample solutions and getting feedback from the
objective function, it learns how to generate solution for complex path
condition containing even black-box function which cannot be solved by
current solver.
{Perry et al.}~\cite{Perry2017} {aim to reduce the cost of
solving array constraints in symbolic execution.
To do so, they present a technique to transform the complex
constraints into the simple one while preserving the semantics.}

\subsection{Software Testing with Machine Learning}
\label{sec:rel4}
{Similar to ours, a few existing techniques use machine learning to
improve software testing}~\cite{Godefroid2017,Wang2017,Koroglu2018,Choi2013,Spieker2017,Sant2005,Cha2018}.
{In Continuous Integration, RECTECS}~\cite{Spieker2017} {uses reinforcement learning
to preferentially execute failing test-cases. Likewise,
in Android GUI testing, QBE}~\cite{Koroglu2018} {employs a standard reinforcement learning algorithm (Q-learning)
to increase both activity coverage and the numbed of crashes.
In web application testing, to achieve high statement coverage, Sant et al.}~\cite{Sant2005}
{automatically build statistical models from logged data via machine learning techniques.
In grammar-based fuzzing, Learn\&Fuzz}~\cite{Godefroid2017}
{leverages recurrent neural networks to automatically learn the complex structure of PDF objects,
intending to maximize code coverage.
In a broad sense, our work belongs to this line of research, where we use a learning algorithm to generate search
heuristics of dynamic symbolic execution automatically.}

\subsection{Search-based Software Testing}
Our work can be seen as an instance of the general framework of
search-based software testing/engineering~\cite{5954405,Harman:2012:SSE:2379776.2379787}, where a testing task is formulated as
an optimization problem and solved by using a meta-heuristic algorithm
(e.g., genetic algorithm). In this work, we formulated the problem of
generating search heuristics of dynamic symbolic execution as an
optimization problem and presented an effective algorithm to solve
it. To our knowledge, this is a novel application from the search-based
software testing perspective.


%% file: conclusion.tex
\section{Conclusion} 

The difficulty of manually crafting good search heuristics has been a
major open challenge in dynamic symbolic execution. In this paper, we
proposed to address this challenge by automatically learning search
heuristics.  Given a program under test, our technique generates a
search heuristic by using a parametric search heuristic and an
optimization algorithm that searches for good parameter values. For
two approaches to dynamic symbolic execution, namely concolic testing
and execution-generated testing, we have shown that 
automatically-generated search heuristics are likely to outperform existing
hand-tuned heuristics, greatly improving the effectiveness of dynamic
symbolic execution.  We hope that our technique can supplant the
laborious and less rewarding task of manually tuning search heuristics
of dynamic symbolic execution.